\documentclass[aps,reprint,nofootinbib,nobibnotes,notitlepage,superscriptaddress,onecolumn,prd,
 amsmath,amssymb
]{revtex4-2}

\usepackage[caption=false]{subfig}
\usepackage{braket}
\usepackage{float}
\usepackage{lipsum}
\usepackage{graphicx}
\usepackage{dcolumn}
\usepackage{bm}
\usepackage{natbib}
\usepackage{hyperref}
\hypersetup{
	colorlinks = true,
    linkcolor = Red,
    urlcolor  = Red,
    citecolor = Red
}
\usepackage[skip=4pt plus1pt, indent=10pt]{parskip}

\usepackage{microtype}

\usepackage{verbatim}
\usepackage[amssymb]{SIunits}
\usepackage{tabularx}
\usepackage[dvipsnames]{xcolor}
\usepackage{wasysym}

\usepackage{tikz}
\usepackage[compat=1.1.0]{tikz-feynman}
\newcommand{\resub}[1]{{#1}}

\def\be{\begin{equation}}
\def\ee{\end{equation}}

\usepackage{color}
\definecolor{darkgreen}{RGB}{0,120,0}

\definecolor{darkgreen}{RGB}{0,120,0}

\newcommand{\av}[1]{\left\langle{#1}\right\rangle}

\newcommand{\vl}{\vec l}
\newcommand{\vL}{\vec L}

\newcommand{\Si}{\mathsf{S}^{-1}}
\newcommand{\F}{\mathcal{F}}
\newcommand{\G}{\mathcal{G}}

\newcommand{\Ai}{\mathsf{A}^{-1}}

\newcommand{\Ci}{\mathsf{C}^{-1}}

\newcommand{\hn}{\hat{\vec n}}

\newcommand{\tj}[6]{\begin{pmatrix} {#1} & {#2} & {#3}\\ {#4} & {#5} & {#6}\end{pmatrix}}

\newcommand{\btheta}{\boldsymbol{\theta}}
\renewcommand{\P}{\mathcal{P}}
\newcommand{\gnldotdot}{g_{\rm NL}^{\dot{\sigma}^4}}

\newcommand{\fnl}{f_{\rm NL}^{\rm loc}}
\newcommand{\gnl}{g_{\rm NL}^{\rm loc}}
\newcommand{\taunl}{\tau_{\rm NL}^{\rm loc}}

\DeclareSymbolFont{toneletters}{T1}{\familydefault}{m}{it}
\SetSymbolFont{toneletters}{bold}{T1}{\familydefault}{bx}{it}

\DeclareMathSymbol\edth{\mathord}{toneletters}{"F0}

\usepackage{adjustbox}
\usepackage{dashbox}

\def\beq{\begin{eqnarray}}
\def\eeq{\end{eqnarray}}

\let\vec\mathbf

\usepackage{empheq}

\def\mnras{\rm{MNRAS}}

\def\aap{\rm{A\&A}}

\defcitealias{Philcox4pt2}{Paper 2}
\defcitealias{Philcox4pt3}{Paper 3}

\usepackage{xspace}

\newcommand{\polyspec}{\textsc{PolySpec}\xspace}

\begin{document}

\title{The ISW-Lensing Bispectrum \& Trispectrum}
\setlength{\parskip}{0pt}

\author{Oliver~H.\,E.~Philcox}
\email{ohep2@cantab.ac.uk}
\affiliation{Simons Society of Fellows, Simons Foundation, New York, NY 10010, USA}
\affiliation{Center for Theoretical Physics, Columbia University, New York, NY 10027, USA}
\affiliation{Department of Physics,
Stanford University, Stanford, CA 94305, USA}
\author{J.\,Colin~Hill}
\affiliation{Center for Theoretical Physics, Columbia University, New York, NY 10027, USA}
\email{jch2200@columbia.edu}

\begin{abstract} 
    \noindent Due to the integrated Sachs-Wolfe (ISW) effect, cosmic microwave background (CMB) temperature and polarization fluctuations are correlated with the gravitational lensing potential. Famously, this induces a CMB three-point function, whose shape can be used to constrain dark energy and modifications to gravity. An analogous effect occurs at higher-order, producing an ISW-lensing \textit{trispectrum} whose amplitude is hitherto unconstrained. We present a detailed discussion of this effect, and define minimum-variance estimators for the ISW-lensing three- and four-point functions. These are implemented within the \href{https://github.com/oliverphilcox/PolySpec}{\textsc{PolySpec}} code, and bear strong similarities to the quadratic estimators used in lensing analyses. Applying these tools to \textit{Planck}, we obtain strong detections of the bispectrum amplitude (consistent with previous works), but find only weak constraints on the trispectrum, due to a strong cancellation between the various ISW-induced contributions. We additionally forecast the constraints from future datasets, finding that (a) simple estimators for the ISW-lensing bispectrum will be severely limited by non-Gaussian modifications to the covariance, and (b) the ISW-lensing trispectrum will be very challenging to detect even with high-resolution future experiments. We finally consider the induced bias on primordial non-Gaussianity amplitudes (and lensing itself), which we show to be large for the bispectrum (as expected) but negligible for the trispectrum.
\end{abstract}

\maketitle

\section{Introduction \& Motivation}
\noindent Cosmic Microwave Background (CMB) photons have undergone an epic journey to reach us today. On their pathway from the surface of last scattering to the redshift-zero observer, they interact with the late Universe in a number of ways. Notable examples include scattering of photons from free electrons (the Sunyaev-Zel'dovich effect \citep{Sunyaev:1980vz}), energy change due to time-varying gravitational potentials (the Integrated Sachs-Wolfe (ISW) effect \citep{Sachs:1967er,Rees:1968zza}) and geodesic deviation imprinted by large-scale distribution of matter (gravitational lensing \citep[e.g.,][]{Lewis:2006fu}). These impart both spectral and/or spatial distortions to the CMB fluctuations, acting both as a late-time signal and an early-time contaminant.

Interactions of CMB photons with large-scale structure generically leads to non-Gaussianity in the observed distributions. As emphasized in a number of previous works \citep[e.g.,][]{Hill:2018ypf,Coulton:2022wln,2011MNRAS.417....2S,Hanson:2009gu,Lewis:2011fk,Junk:2012qt,Serra:2008wc,Curto:2013hi,Curto:2014bna,Kim:2013nea,Calabrese:2009bu}, this can lead to non-trivial biases in measurements of primordial non-Gaussianity parameters, which encode novel physics in inflation, such as particle interactions and the cosmological collider. The degree of bias depends on a number of factors, including the type of contaminant, the resolution of the experiment, and the overlap between primordial and late-time templates; if it is found to be significant, it must be carefully subtracted to avoid spurious detections of primordial phenomena.

Perhaps the most well-studied source of late-time non-Gaussianity is CMB lensing. This results in a remapping of the CMB temperature field with $T\to \tilde{T}\equiv T+\nabla T\cdot\nabla\phi$ at leading order, where $\phi$ is the lensing potential, which involves a line-of-sight integral over the matter distribution. Famously, lensing induces a four-point function with the schematic form:
\beq\label{eq: lensing-4pt-schem}
    \av{\tilde{T}\tilde{T}TT}_c\sim \av{T\nabla T}^2\av{\nabla\phi\nabla\phi};
\eeq
this encodes the power spectrum of the lensing potential, and thus the late-time distribution of dark matter. Starting from \citep{Smith:2007rg}, this has been used to measure the lensing power spectrum to progressively higher signal-to-noise, with $>40\sigma$ detections reported in both ACT \citep{ACT:2023dou} and \textit{Planck} \citep{Philcox4pt3,Carron:2022eyg} analyses. 

Further non-Gaussianity can be formed if the lensing potential is correlated to the unlensed temperature (or polarization) field. Such a correlation naturally arises through the ISW effect, which is sourced by the same gravitational potentials which cause geodesic deviations. At leading-order, this induces a three-point function, known a the ISW-lensing bispectrum \citep[e.g.,][]{Goldberg:1999xm,Boubekeur:2009uk,Lewis:2011fk}:
\beq
    \av{\tilde{T}TT}_c \sim \av{T\nabla T}\av{T\nabla\phi};
\eeq
this is an important contaminant to local primordial non-Gaussianity studies 
\citep{Hill:2018ypf,2011MNRAS.417....2S,Hanson:2009gu,Lewis:2011fk,Junk:2012qt,Planck:2015zfm,Planck:2019kim,Kim:2013nea,Calabrese:2009bu}. Additional correlations can also be present, including through thermal and kinematic Sunyaev-Zel'dovich effects, unresolved point sources and the cosmic infrared background \citep{Hill:2018ypf,Coulton:2022wln,Serra:2008wc,Curto:2013hi,Curto:2014bna}, though many can be mitigated using frequency-based cleaning. At second-order, correlations between the ISW effect and CMB lensing induce a four-point function:
\beq\label{eq: isw-lensing-4pt-schem}
    \av{\tilde{T}\tilde{T}TT}_c\sim 
    \av{T\nabla \phi}^2\av{\nabla T\nabla T};
\eeq
this is analogous to the lensing trispectrum of \eqref{eq: lensing-4pt-schem}, except that the two lensed fields `exchange' an unlensed field instead of the lensing potential.\footnote{In the language of quadratic estimators, we are building an estimator for the unlensed field, rather than for the lensing potential.} As discussed below, this trispectrum recieves a number of additional contributions, including from higher-order terms in the lensing expansion (e.g., the $\nabla\nabla T\nabla \phi \nabla\phi$ contribution to $\tilde{T}$).

Whilst the ISW-lensing bispectrum has been extensively studied both theoretically \citep[e.g.,][]{Boubekeur:2009uk,Giannantonio:2013uqa,Hill:2018ypf,Lewis:2011fk,Mangilli:2013sxa,Goldberg:1999xm,Kim:2013nea,Foreman:2018lci} and observationally \citep[e.g.,][]{Carron:2022eum,Carron:2022eyg,Planck:2013mth,Planck:2015mym,Mangilli:2013sxa} (including its use as a probe of modified gravity and cosmic strings \citep{Hu:2012td,Yamauchi:2013pna,Munshi:2014tua,Kable:2021yws,Jain:2007yk,Giannantonio:2013kqa,Calabrese:2009tt,Chudaykin:2025gdn}), there has been little-to-no prior discussion of the ISW-lensing trispectrum. In this work, we perform a detailed investigation of the latter effect, seeking to answer the following questions: (1) is the signal detectable? (2) does it bias measurements of CMB lensing? (3) does it bias measurements of primordial trispectrum amplitudes, such as $g_{\rm NL}$ and $\tau_{\rm NL}$? To do this, we will build optimal estimators for the signals of interest and implement them in the public code \href{https://github.com/oliverphilcox/PolySpec}{\textsc{PolySpec}}, which allow for efficient Fisher forecasts across a wide range of scales. As an initial exercise, we will derive estimators for ISW-lensing bispectra, which we will apply to the latest \textit{Planck} temperature and polarization dataset, finding analogous results to previous studies \citep[e.g.,][]{Lewis:2011fk,Carron:2022eum}. 

\vskip 4pt
The remainder of this paper is as follows. We begin with a pedagogical discussion of ISW-lensing cross-correlations in \S\ref{sec: flat-sky}, working in the flat-sky limit for interpretability. In \S\ref{sec: full-sky}\,\&\,\S\ref{sec: estimators}, we present the full-sky correlators and introduce minimum-variance estimators for each effect, facilitating their efficient measurement and forecasting. In \S\ref{sec: current-data}, we constrain ISW-lensing bispectra and trispectra using \textit{Planck} data, before presenting forecasts for future idealized experiments in \S\ref{sec: forecasts}. We conclude with a discussion in \S\ref{sec: summary}. In Appendix \ref{app: post-born}, we discuss higher-order effects ignored in the main text, whilst Appendix \ref{app: pol} discusses the extension to polarization, and we list various functions used to compute the normalization matrices in Appendix \ref{app: Q-derivs}. Throughout this work, we assume the \textit{Planck} 2018 cosmology: $\{h = 0.6732, \omega_b = 0.02238, \omega_c = 0.1201, \tau_{\rm reio} = 0.05431, n_s = 0.9660, A_s =2.101\times 10^{-9}, \sum m_\nu = 0.06\,\mathrm{eV}\}$ with a single massive neutrino \citep{2020A&A...641A...6P}.

\section{Warm-Up: Flat Sky Correlators}\label{sec: flat-sky}
\noindent To build intuition for the main results of this work, we first discuss the flat-sky limits of the various lensing- and ISW-induced correlation functions, restricting our attention to temperature anisotropies for simplicity. Throughout, we assume the Born approximation (evaluating the lensing distortion on unperturbed geodesics) and neglect any non-Gaussianity in the lensing potential;\footnote{We do include non-linear corrections to the lensing power spectrum, however, using \textsc{halofit} \citep{Takahashi:2012em}.} as discussed in Appendix \ref{app: post-born}, these effects source only small corrections to the ISW-lensing statistics.

\vskip8pt
\paragraph{The Lensing \& Integrated Sachs-Wolfe Effects}
\noindent Time variations in the gravitational potential, $\Phi$, induce changes in the temperature of CMB photons. On the full sky, this sources the perturbation 
\beq
    \left(\frac{\Delta T_{\rm ISW}}{T_{\rm CMB}}\right)(\hn) = -\frac{2}{c^2}\int_0^{\chi_\star}d\chi\,\partial_\chi\Phi(\chi\hn,\chi)
\eeq
at observation angle $\hn$, where we integrate over all possible sources from last scattering (at conformal distance $\chi=\chi_\star$) to today ($\chi=0$). The gravitational potential $\Phi$ also sources CMB lensing; this is described by the lensing potential, $\phi$, which, for a source at distance $\chi$, is given by
\beq\label{eq: lensing-potential}
    \phi(\hn,\chi_s) = -\frac{2}{c^2}\int_0^{\chi_s}d\chi'\frac{\chi_s-\chi'}{\chi_s\chi'}\Phi(\chi'\hn,\chi');
\eeq
this involves an integral over all distances from the source to the observer. Since the two quantities involve over similar redshift ranges, they are correlated, leading to a temperature-lensing cross-spectrum. In the Limber approximation, this has the leading-order form:
\beq\label{eq: ISW-lensing-Cl}
    C_\ell^{T\phi}(\chi_s) \equiv \av{T_{\ell m}\phi_{\ell m}^*(\chi_s)} = \frac{2}{c^4}\int_0^{\chi_s}\frac{d\chi'}{\chi'^2}\frac{\chi_s-\chi'}{\chi_s\chi'}\frac{\partial P_\Phi}{\partial \chi'}(\ell+1/2)/\chi',\chi')
\eeq
\citep[e.g.,][]{Hill:2018ypf,Goldberg:1999xm}, where $P_{\Phi}$ is the power spectrum of the Weyl potential. Typically, we consider lensing of the primary perturbations, whence $\chi_s=\chi_\star$, though below we will also consider lensing of late-time sources, with $\chi_s\ll \chi_\star$. The $T\phi$ cross-correlation traces physics in the dark-energy-dominated regime; moreover, it peaks on very large scales. 
The dominance of low $\ell$ leads to the particular geometric properties of the ISW-lensing correlators which, as described below, are central to this work.

\vskip8pt
\paragraph{Lensing Non-Gaussianity}
\noindent Gravitational lensing induces a remapping of the temperature perturbations from the unlensed position $\btheta$ to the lensed position $\btheta+\nabla\phi(\btheta)$ (see \citep{Lewis:2006fu} for a review). For a source at conformal distance $\chi_s$ with unlensed temperature $T(\btheta,\chi_s)$ (with $\chi_s=\chi_\star$ for the primary perturbations), the lensed field has the flat-sky limit
\beq\label{eq: lensing-flat-sky}
    \tilde{T}(\btheta,\chi_s) &\equiv& T(\btheta+\nabla\phi(\btheta,\chi_s),\chi_s)=T(\btheta,\chi_s)+\nabla_iT(\btheta,\chi_s)\nabla^i\phi(\btheta,\chi_s)\\\nonumber
    &&\,+\,\frac{1}{2}\nabla_i\nabla_j T(\btheta,\chi_s)\nabla^i\phi(\btheta,\chi_s)\nabla^j\phi(\btheta,\chi_s)+\cdots,
\eeq
where $\nabla_i\equiv\partial_{\theta_i}$ are derivatives on the two-dimensional lensing plane. Here, $\phi(\btheta,\chi_s)$ is the lensing potential for a source at distance $\chi_s$: this integrates over the gravitational potential for all $\chi\in[0,\chi_s']$, as in \eqref{eq: lensing-potential}.\footnote{Usually, one assumes that lensing acts only on the primary CMB field, \textit{i.e.}\ we fix $\chi_s=\chi_\star$. This assumption is violated for certain ISW-lensing cross-correlations, as discussed below. We thank Anthony Challinor for pointing this out.} This is typically expressed in Fourier-space, with $X(\vl,\chi_s) \equiv \int d\btheta\, e^{-i\vl\cdot\btheta}X(\btheta,\chi_s)$:
\beq
    \tilde{T}(\vl,\chi_s) &=& T(\vl,\chi_s) - \int_{\vL}[\vL\cdot(\vl-\vL)] T(\vl-\vL,\chi_s)\phi(\vL,\chi_s)\\\nonumber
    &&\,+\,\frac{1}{2}\int_{\vL\vL'}[\vL\cdot(\vl-\vL-\vL')][\vL'\cdot(\vl-\vL-\vL')]T(\vl-\vL-\vL',\chi_s)\phi(\vL,\chi_s)\phi(\vL',\chi_s)+\cdots,
\eeq
denoting $\int_{\vL} \equiv (2\pi)^{-2}\int d\vL$. Finally, we can integrate over all possible sources, including the primary CMB and any late-time contributions (such as the ISW effect): 
\beq\label{eq: T-lens-flat}
    \tilde{T}(\vl) &=& T(\vl) - \int_{\vL}[\vL\cdot(\vl-\vL)] \int_0^{\chi_\star}d\chi\,T(\vl-\vL,\chi)\phi(\vL,\chi)\\\nonumber
    &&\,+\,\frac{1}{2}\int_{\vL\vL'}[\vL\cdot(\vl-\vL-\vL')][\vL'\cdot(\vl-\vL-\vL')]\int_0^{\chi_\star}d\chi_s\,T(\vl-\vL-\vL',\chi)\phi(\vL,\chi)\phi(\vL',\chi)+\cdots,
\eeq
where, by definition, $T(\vl)\equiv\int_0^{\chi_\star}d\chi\,T(\vl,\chi)$.

\vskip8pt
\paragraph{Bispectra}
\noindent To form the ISW-lensing bispectrum, we contract the first-order-in-$\phi$ term in \eqref{eq: T-lens-flat} with two unlensed fields, finding the flat-sky form\footnote{Here and henceforth, $\av{\cdots}_c'$ represents the connected correlator, stripping the momentum-conserving delta function.}
\beq\label{eq: bispec-flat}
    \av{\tilde{T}(\vl_1)\tilde{T}(\vl_2)\tilde{T}(\vl_3)}'_c &=& -[\vl_2\cdot\vl_3]\int_0^{\chi_\star}d\chi\,\av{T(\vl_2)T^*(\vl_2,\chi)}\av{T(\vl_3)\phi^*(\vl_3,\chi)} + \text{5 perms.}\\\nonumber
    &\equiv& -[\vl_2\cdot\vl_3]\int_0^{\chi_\star}d\chi\,C^{TT}_{l_2}(\chi)C^{T\phi}_{l_3}(\chi) + \text{5 perms.},
\eeq
adopting the asymmetric definition $\av{X(\vl)Y^*(\vl',\chi)} \equiv (2\pi)^2\delta_{\rm D}(\vl-\vl')C_l^{XY}(\chi)$. This involves the cross-correlation of the full temperature field with that sourced by temperature perturbations at $\chi$ (e.g., the ISW contribution at low $\chi$ or the primary piece at $\chi\approx \chi_\star$), as well as the cross-correlation of the full temperature field with the lensing potential for a source at distance $\chi$. The first term is dominated by the primary anisotropies at $\chi\approx \chi_\star$; in this limit, we find the canonical ISW-lensing bispectrum:
\beq\label{eq: bispec-flat}
    \av{\tilde{T}(\vl_1)\tilde{T}(\vl_2)\tilde{T}(\vl_3)}'_c = -[\vl_2\cdot\vl_3]C_{l_2}^{TT}C_{l_3}^{T\phi} + \text{5 perms.},
\eeq
defining $C_l^{T\phi}(\chi_\star)\equiv C_l^{T\phi}$.
Whilst the above derivation would suggest that the power spectra entering \eqref{eq: bispec-flat} should be unlensed, a more accurate model is obtained by using lensed power spectra (or, more accurately still, derivative spectra, such as $C_l^{T\nabla T}$), as discussed in \citep{Lewis:2011fk}. Since $C_l^{T\phi}$ is negligible on small-scales, this peaks in squeezed configurations,\footnote{Note that the leading order in the squeezed limit vanishes after symmetrization, e.g., setting $(\vl_1,\vl_2,\vl_3) = (\vl,-\vl-\vL,\vL)$ with $l\gg L$, \eqref{eq: bispec-flat} asymptotes to $\left(1+(\partial\log C_{l}^{TT}/\partial\log l)\cos^2\theta\right)L^2C_l^{TT}C_{L}^{T\phi}$, where $\theta$ is the angle between $\vl$ and $\vL$.} and is thus an important contaminant to multi-field inflation signals such as $f_{\rm NL}^{\rm loc}$ \citep{Goldberg:1999xm,Boubekeur:2009uk,Lewis:2011fk,Hill:2018ypf,2011MNRAS.417....2S,Hanson:2009gu,Lewis:2011fk,Junk:2012qt,Planck:2015zfm,Planck:2019kim,Kim:2013nea,Calabrese:2009bu}.

\vskip8pt
\paragraph{Trispectra}
\noindent There are two ways to create a lensing trispectrum: (1) correlating two first-order terms in \eqref{eq: T-lens-flat} with two unlensed fields; (2) correlating a second-order term in \eqref{eq: T-lens-flat} with three unlensed fields. The first leads to the following contributions:
\beq\label{eq: flat-sky-tspec1-chi}
    \av{\tilde{T}(\vl_1)\tilde{T}(\vl_2)\tilde{T}(\vl_3)\tilde{T}(\vl_4)}'_c &\supset& 
    -\int_0^{\chi_\star} d\chi\int_0^{\chi_\star} d\chi'\int_{\vL}(2\pi)^2\delta_{\rm D}(\vl_1+\vl_3-\vL)[\vL\cdot\vl_3] [\vL\cdot\vl_4]\\\nonumber
    &&\,\times\,\bigg[C^{TT}_{l_3}(\chi)C^{TT}_{l_4}(\chi')C^{\phi\phi}_L(\chi,\chi')+C^{T\phi}_{l_3}(\chi)C^{T\phi}_{l_4}(\chi')C^{TT}_L(\chi,\chi')\\\nonumber
    &&\qquad\,+\,C^{TT}_{l_3}(\chi)C^{T\phi}_{l_4}(\chi')C^{T\phi}_L(\chi',\chi)+C^{T\phi}_{l_3}(\chi)C^{TT}_{l_4}(\chi')C^{T\phi}_L(\chi,\chi')\bigg]+\text{11 perms.},
\eeq
with $\av{X(\vl)Y^*(\vl',\chi)} = (2\pi)^2\delta_{\rm D}(\vl-\vl')C_l^{XY}(\chi)$ and $\av{X(\vl,\chi)Y^*(\vl',\chi')} = (2\pi)^2\delta_{\rm D}(\vl-\vl')C_l^{XY}(\chi,\chi')$, whilst the second sources
\beq\label{eq: flat-sky-tspec2-chi}
    \av{\tilde{T}(\vl_1)\tilde{T}(\vl_2)\tilde{T}(\vl_3)\tilde{T}(\vl_4)}'_c &\supset& [\vl_2\cdot\vl_3][\vl_2\cdot\vl_4]\int_0^{\chi_\star}d\chi\,C^{TT}_{l_2}(\chi)C^{T\phi}_{l_3}(\chi)C^{T\phi}_{l_4}(\chi)+\text{11 perms.}
\eeq
Analogously to the ISW-lensing bispectrum, the power spectra entering \eqref{eq: flat-sky-tspec1-chi}\,\&\,\eqref{eq: flat-sky-tspec2-chi} should be lensed.\footnote{This can be derived through a similar approach to \citep{Lewis:2011fk}, working in the limit of $l_1,l_2\ll l_3,l_4$, which dominates the ISW-lensing signal.}. Notably, the $\chi$ integrals involving $C_{l}^{TT}(\chi)$ are strongly dominated by $\chi\approx\chi_\star$ (as for the bispectrum); similarly those involving $C_L^{TT}(\chi,\chi')$ have $\chi\approx\chi'\approx\chi_\star$. In this limit (which we shall assume throughout this work), the trispectrum can be simplified: 
\beq\label{eq: flat-sky-tspec}
    \av{\tilde{T}(\vl_1)\tilde{T}(\vl_2)\tilde{T}(\vl_3)\tilde{T}(\vl_4)}'_c &\supset& 
    -\int_{\vL}(2\pi)^2\delta_{\rm D}(\vl_1+\vl_3-\vL)[\vL\cdot\vl_3] [\vL\cdot\vl_4]\\\nonumber
    &&\,\times\,\bigg[C^{TT}_{l_3}C^{TT}_{l_4}C^{\phi\phi}_L+C^{T\phi}_{l_3}C^{T\phi}_{l_4}C^{TT}_L\\\nonumber
    &&\qquad\,+\,\int_0^{\chi_\star}d\chi\left(C^{TT}_{l_3}C^{T\phi}_{l_4}(\chi)C^{\phi T}_L(\chi)+C^{T\phi}_{l_3}(\chi)C^{TT}_{l_4}C^{\phi T}_L(\chi)\right)\bigg]+\text{11 perms.}\\\nonumber
    &&\,+\,[\vl_2\cdot\vl_3][\vl_2\cdot\vl_4]C^{TT}_{l_2}C^{T\phi}_{l_3}C^{T\phi}_{l_4}+\text{11 perms.}
\eeq
where $C_l^{\phi T}(\chi)$ correlates the temperature sourced at $\chi$ with the full potential $\phi\equiv \phi(\chi_\star)$\footnote{This should not be confused with $C_l^{T\phi}(\chi)$, which correlates the full temperature $T$ with the lensing potential for a source at distance $\chi$.}; this vanishes for large $\chi$, implying that the remaining $\chi$ integral is sourced only by low-redshift contributions, \textit{i.e.}\ the lensing of late-time effects.

\begin{figure}
    \centering
    \includegraphics[width=\linewidth]{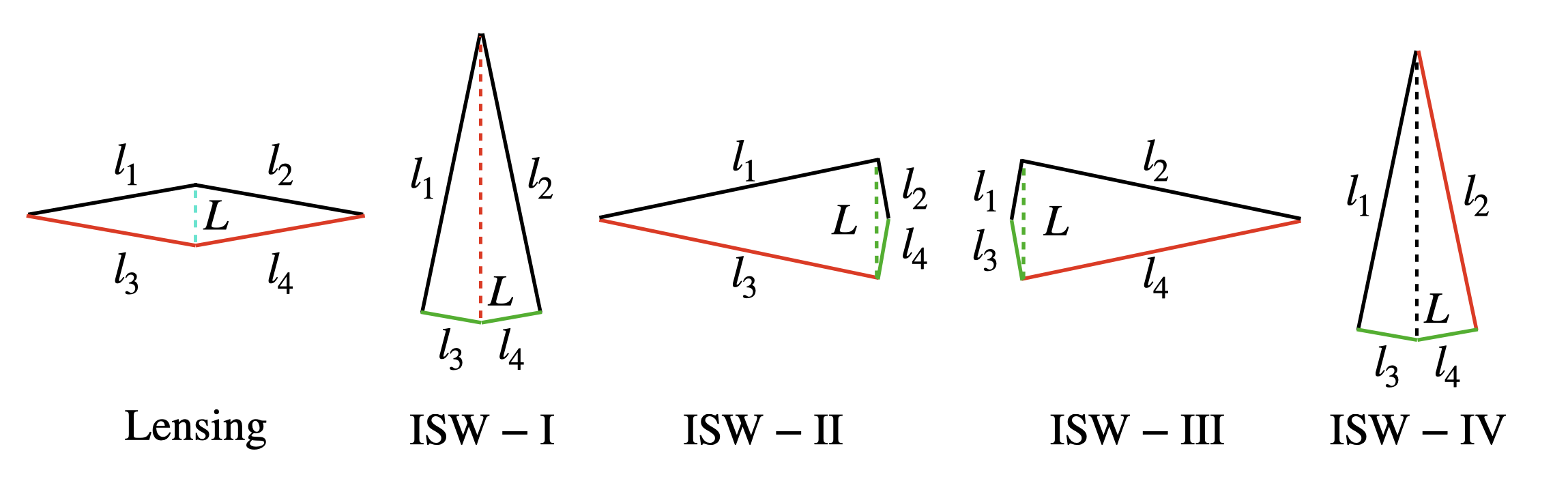}
    \caption{Schematic of the lensing and ISW-lensing trispectra. For each shape (whose flat-sky form is given in \S\ref{sec: flat-sky}), we show the type of harmonic-space quadrilateral that dominates the signal-to-noise, for example, displaying a collapsed configuration for lensing. Colors indicate the type of correlators, with red, blue, and green lines representing $C_l^{TT},C_l^{\phi\phi}$ and $C_l^{T\phi}$ respectively. The full ISW-lensing trispectrum is the sum of four kinds of contribution, each of which peaks in very different regimes to the lensing configuration. In practice, we find that ISW-II and ISW-III are small (being generated by lensing of the ISW effect), and that ISW-I and ISW-IV almost cancel, greatly reducing the signal-to-noise.}
    \label{fig: cartoon}
\end{figure}

There are several types of contribution to \eqref{eq: flat-sky-tspec}. The first term, proportional to $C_{l_3}^{TT}C_{l_4}^{TT}C_{L}^{\phi\phi}$, is the usual CMB lensing trispectrum \citep[e.g.,][]{Lewis:2006fu}, involving two external temperature power spectra and internal lensing power spectrum. This typically peaks at large $l_3,l_4$ and low $L$, \textit{i.e.}\ in collapsed configurations, as shown in Fig.\,\ref{fig: cartoon}. The second through fifth terms (hereafter denoted ISW-I through ISW-IV) are sourced by cross-correlations of $\phi$ with $T$ due to the ISW effect (or any other late-time contamination \citep[cf.][]{Hill:2018ypf}). These have the following properties, which are shown graphically in Fig.\,\ref{fig: cartoon}:
\begin{itemize}
    \item \textbf{ISW-I}: This involves two external ISW-lensing spectra and an internal primary spectrum. Since $C_l^{T\phi}$ is dominated by large-scales, this peaks in the doubly-squeezed limit, with small $l_3,l_4$ but large $l_1,l_2,L=|\vl_1+\vl_3|$ (and permutations thereof).
    \item \textbf{ISW-II}: This involves an external primary spectrum and both internal and external ISW-lensing spectra. This is dominated by small $l_2,l_4,L$ and large $l_1,l_3$. Physically, this is sourced by the lensing of the ISW, rather than the lensing of the primary CMB.
    \item \textbf{ISW-III}: This is equivalent to ISW-II after permutations.
    \item \textbf{ISW-IV}: Unlike the previous `exchange' contributions, this is a `contact' trispectrum, which contains only external legs. This peaks at large $l_1,l_2$ but small $l_3,l_4$, and thus large diagonal momentum $L$.
\end{itemize}

Whilst ISW-I and ISW-IV have different origins, they have similar behavior in the collapsed limit. Setting $l_3\ll l_1,L$ and $l_4\ll l_2,L$, we find the asymptotic results
\beq
    \left.\av{\tilde{T}(\vl_1)\tilde{T}(\vl_2)\tilde{T}(\vl_3)\tilde{T}(\vl_4)}'_c\right|_{\rm ISW-I} &\to& -[\vl_2\cdot\vl_3] [\vl_2\cdot\vl_4]C^{T\phi}_{l_3}C^{T\phi}_{l_4}C^{TT}_{l_2}+\text{11 perms.}\\\nonumber
    \left.\av{\tilde{T}(\vl_1)\tilde{T}(\vl_2)\tilde{T}(\vl_3)\tilde{T}(\vl_4)}'_c\right|_{\rm ISW-IV} &\to& +[\vl_2\cdot\vl_3][\vl_2\cdot\vl_4]C^{T\phi}_{l_3}C^{T\phi}_{l_4}C^{TT}_{l_2}+\text{11 perms.},
\eeq
thus the sum cancels at leading-order. This significantly reduces the signal-to-noise of the quartic ISW-lensing effect and is analogous to the second-order cancellations in the CMB lensing power spectrum \citep{Challinor:2005jy}.

When deriving estimators for CMB lensing, one usually adopts the `quadratic estimator' perspective, whence a pair of lensed temperature (or polarization) fields form an estimator for the lensing potential $\phi$, \textit{i.e.}\ we consider the contraction
\beq\label{eq: quad-lensing}
    \av{\tilde{T}(\vl_1)\tilde{T}(\vl_2)}'_{\rm unlensed} &=& \left[(\vl_1+\vl_2)\cdot\vl_2\right]C_{l_2}^{TT}\phi(\vl_1+\vl_2) + (\vl_1\leftrightarrow\vl_2),
\eeq
averaging over the unlensed CMB fluctuations and noting that the $\chi$ integral is dominated by $\chi\approx \chi_\star$, as before. The lensing power spectrum is obtained by squaring the above expression (and removing a number of biases), \textit{i.e.}\ from the temperature four-point function \citep[e.g.,][]{Carron:2022edh,Hu:2001kj,Okamoto:2003zw}. A similar approach can be used to extract the ISW-lensing correlators; instead of averaging over the unlensed fluctuations, we take the expectation over ISW-lensing correlations, such that
\beq\label{eq: quad-ISW}
    \av{\tilde{T}(\vl_1)\tilde{T}(\vl_2)}'_{\rm ISW} &=& \left[(\vl_1+\vl_2)\cdot\vl_2\right]\int_0^{\chi_\star}d\chi\,C_{l_2}^{T\phi}(\chi)T(\vl_1+\vl_2,\chi) + (\vl_1\leftrightarrow\vl_2);
\eeq
this implies that a pair of lensed fields provides an estimator for the unlensed field $T(\vl_1+\vl_2,\chi)$.\footnote{This bears similarities to the approach of \citep{Foreman:2018lci}, which cross-correlate small-scale temperature and lensing fields to reconstruct the large-scale temperature, which can be used to improve measurements of the ISW effect.} As in the lensing case, we can immediately extract the trispectra: ISW-I is formed by squaring \eqref{eq: quad-ISW}, whilst ISW-II and ISW-III are formed by cross-correlating \eqref{eq: quad-ISW} with \eqref{eq: quad-lensing}. Since the quadratic estimators for $\phi$ and $T$ are dominated by modes in very different regimes, we typically expect the cross-correlation terms to be small.

The above discussion yields a number of conclusions about the ISW-lensing trispectrum, which will we verify in \S\ref{sec: current-data}. Firstly, it is clear that the ISW-lensing trispectrum is a real physical signal that must be present in the observational data. That said, it is unclear whether the effect is large enough to be detected in current or future data given that (a) two of the terms cancel at leading order, and (b) the other two terms arise from cross-correlations of quadratic estimators dominated by very different scales, and are sourced only by lensing of low-redshift anisotropies. Finally, we note that all of the trispectrum components peak in the doubly squeezed regimes (Fig.\,\ref{fig: cartoon}). This is very different to both the lensing shape and many inflationary templates (such as $\gnl$, which is singly-squeezed, and $\taunl$, which is collapsed), thus it seems unlikely that this foreground can pose significant bias to lensing or primordial non-Gaussianity studies, in contrast to its effect on primordial bispectrum estimators.

\section{Full-Sky Correlators}\label{sec: full-sky}
\subsection{Full-Sky Lensing}
\noindent Under the Born approximation (cf.\,Appendix \ref{app: post-born}), gravitational lensing remaps the angular coordinates $\hn\equiv(\theta,\phi)$ to $\hn + \nabla\phi(\hn)$, where $\phi$ is the lensing potential discussed above, and $\nabla$ is a covariant derivative on the two-sphere \citep{Okamoto:2002ik,Okamoto:2003zw,Lewis:2006fu}. For a temperature source at conformal distance $\chi$, this has the perturbative expansion
\beq\label{eq: lensing-full-sky}
    \tilde{T}(\hn,\chi_s) &\to& T(\hn+\nabla\phi(\hn,\chi_s),\chi_s) \equiv T(\hn,\chi_s) + \nabla_iT(\hn,\chi_s)\nabla^i\phi(\hn,\chi_s) + \frac{1}{2}\nabla_i\nabla_jT(\hn,\chi_s)\nabla^i\phi(\hn,\chi_s)\nabla^j\phi(\hn,\chi_s)+\cdots
\eeq
analogous to \eqref{eq: lensing-flat-sky}. Integrating over sources and transforming to spherical-harmonic-space, we find
\beq\label{eq: lensing-full-sky-harmonic}
    \tilde{a}_{\ell m}&=& a_{\ell m} + (-1)^m\sum_{\ell'm'LM}\mathcal{I}^{\ell m}_{\ell'm'LM}\int_0^{\chi_\star}d\chi\,a_{\ell'm'}(\chi)\phi_{LM}(\chi)\\\nonumber
    &&\,+\,\frac{1}{2}(-1)^{m}\sum_{\ell'm'LML'M'}\mathcal{J}^{\ell m}_{\ell'm'LML'M'}\int_0^{\chi_\star}d\chi\,a_{\ell'm'}(\chi)\phi_{LM}(\chi)\phi_{L'M'}(\chi)+\cdots,
\eeq
where $a_{\ell m}(\chi)$ is the harmonic transform of $T(\hn,\chi)$ (with $a_{\ell m}\equiv\int_0^{\chi_\star}d\chi\,a_{\ell m}(\chi)$, as before), and $\mathcal{I}$ and $\mathcal{J}$ are coupling kernels, which can be written in terms of spin-weighted spherical harmonics:\footnote{Our $\mathcal{I}$ is equivalent to which is equal to $(-1)^mI^{mm''m'}_{\ell\ell''\ell'}$ in the notation of \citep{Okamoto:2003zw}. Our $\mathcal{J}$ is the spin-zero limit of the full form presented in Appendix \ref{app: pol}.}
\beq\label{eq: quadratic-kernel}
    \mathcal{I}^{\ell m}_{\ell'm'LM} &\equiv& (-1)^m\int d\hn\,Y^*_{\ell m}(\hn)\nabla_i Y_{\ell'm'}(\hn)\nabla_i Y_{LM}(\hn)\\\nonumber
    &=&-\frac{1}{2}\sqrt{L(L+1)}\sqrt{\ell'(\ell'+1)}\sum_{\lambda=\pm1}(-1)^m\int d\hn\,Y^*_{\ell m}(\hn){}_{\lambda}Y_{\ell'm'}(\hn){}_{-\lambda}Y_{LM}(\hn)
\eeq
and
\beq\label{eq: cubic-kernel}
    \mathcal{J}^{\ell m}_{\ell'm'LML'M'} &\equiv& (-1)^{m}\int d\hn\,Y^*_{\ell m}(\hn)\nabla_i\nabla_jY_{\ell'm'}(\hn)\nabla_iY_{LM}(\hn)\nabla_jY_{L'M'}(\hn)\\\nonumber
    &=&(-1)^{m}\frac{\sqrt{\ell'(\ell'+1)L(L+1)L'(L'+1)}}{4}\sum_{\lambda=\pm1}\int d\hn\,Y^*_{\ell m}(\hn)\\\nonumber
    &&\,\times\,\bigg[\sqrt{\ell'(\ell'+1)}Y_{\ell'm'}(\hn){}_{\lambda}Y_{LM}(\hn)+\sqrt{(\ell'-1)(\ell'+2)}{}_{2\lambda}Y_{\ell'm'}(\hn){}_{-\lambda}Y_{LM}(\hn)\bigg]{}_{-\lambda}Y_{L'M'}(\hn)
\eeq
using \citep{Okamoto:2003zw}. These can be expressed in terms of Wigner $3j$ symbols, though the above forms are more useful in this work. Notably, $\mathcal{I}$ is symmetric under $(\ell',m')\leftrightarrow(L,M)$, \textit{i.e.}\ it treats both the unlensed and potential fields equivalently.

As discussed in \citep{Okamoto:2003zw}, similar forms can be derived for the spin-two polarization tensor, $\P_{ij}(\hn)$, and thus the CMB $E$- and $B$-modes. These take a somewhat more complex (and asymmetric) form due to the addition of polarization indices, and are presented in Appendix \ref{app: pol}.

\subsection{Correlation Functions}\label{subsec: correlators}
\noindent The (ISW-)lensing bispectra and trispectra can be immediately extracted from \eqref{eq: lensing-full-sky-harmonic}. At leading-order, the temperature bispectrum is (dropping tildes for clarity)
\beq\label{eq: bis-full-sky-T}
    \av{a_{\ell_1m_1}a_{\ell_2m_2}a_{\ell_3m_3}}_c&\supset&\mathcal{I}^{\ell_1m_1}_{\ell_2(-m_2)\ell_3(-m_3)}C_{\ell_2}^{TT}C_{\ell_3}^{T\phi}+\text{5 perms.},
\eeq
matching \citep{Lewis:2011fk}, where $C_\ell^{T\phi}$ is the cross-spectrum induced by ISW-lensing correlations as in \eqref{eq: ISW-lensing-Cl}. Here, we have taken the $\chi\approx\chi_\star$ limit as in \S\ref{sec: flat-sky}. This can be recast in terms of the reduced bispectrum, $b_{\ell_1\ell_2\ell_3}$, for even $\ell_1+\ell_2+\ell_3$:
\beq
    \av{a_{\ell_1m_1}a_{\ell_2m_2}a_{\ell_3m_3}}_c &\equiv& 
    \G^{\ell_1\ell_2\ell_3}_{m_1m_2m_3}b_{\ell_1\ell_2\ell_3}\\\nonumber
    b_{\ell_1\ell_2\ell_3}&\supset&\frac{1}{2}\left[\ell_2(\ell_2+1)+\ell_3(\ell_3+1)-\ell_1(\ell_1+1)\right]C_{\ell_2}^{TT}C_{\ell_3}^{T\phi}+\text{5 perms.},
\eeq
inserting the Gaunt symbol ($\G$), and simplifying. This peaks in the squeezed regime, where $\ell_3\ll \ell_1,\ell_2$ (and permutations). As shown in Appendix \ref{app: pol}, the polarization bispectrum is somewhat more complex, and is non-zero for both even and odd $\ell_1+\ell_2+\ell_3$.

The full-sky trispectrum is analogous to the flat-sky case discussed in \S\ref{sec: flat-sky}. First-order lensing sources four contributions, as depicted in Fig.\,\ref{fig: cartoon}:
\beq\label{eq: tris-ex-full-sky-T}
    \av{a_{\ell_1m_1}a_{\ell_2m_2}a_{\ell_3m_3}a_{\ell_4m_4}}_c &\supset& \sum_{LM}(-1)^{M}\mathcal{I}^{\ell_1m_1}_{LM\ell_3(-m_3)}\mathcal{I}^{\ell_2m_2}_{L(-M)\ell_4(-m_4)}\\\nonumber
    &&\,\times\,\bigg[C_{\ell_3}^{TT}C_{\ell_4}^{TT}C_L^{\phi\phi}+C_{\ell_3}^{T\phi}C_{\ell_4}^{T\phi}C_L^{TT}\\\nonumber
    &&\qquad\,+\,\int_0^{\chi_\star} d\chi\,\left(C_{\ell_3}^{TT}C_{\ell_4}^{T\phi}(\chi)C_L^{\phi T}(\chi)+C_{\ell_3}^{T\phi}(\chi)C_{\ell_4}^{TT}C_L^{\phi T}(\chi)\right)\bigg] + \text{11 perms.},
\eeq
where the first is the usual CMB lensing trispectrum, the second involves the exchange of an unlensed field (ISW-I), and the final two are mixed terms (ISW-II \& ISW-III), which depend explicitly on low-redshift contributions. Similarly to the bispectrum, this can be expressed in the reduced form defined in \citep{Regan:2010cn}:
\beq\label{eq: tris-con-full-sky-T}
    &&\av{a_{\ell_1m_1}a_{\ell_2m_2}a_{\ell_3m_3}a_{\ell_4m_4}}_c \equiv \sum_M(-1)^M\G^{\ell_1\ell_2L}_{m_1m_2(-M)}\G^{\ell_3\ell_4L}_{m_3m_4M}t^{\ell_1\ell_2}_{\ell_3\ell_4}(L)+\text{11 perms.}\\\nonumber
    t^{\ell_1\ell_2}_{\ell_3\ell_4}(L) &\supset& \frac{1}{4}\left[L(L+1)+\ell_2(\ell_2+1)-\ell_1(\ell_1+1)\right]\left[L(L+1)+\ell_4(\ell_4+1)-\ell_3(\ell_3+1)\right]\\\nonumber
    &&\,\times\,\bigg[C_{\ell_2}^{TT}C_{\ell_4}^{TT}C_L^{\phi\phi}+C_{\ell_2}^{T\phi}C_{\ell_4}^{T\phi}C_L^{TT}+\int_0^{\chi_\star} d\chi\,\left(C_{\ell_2}^{TT}(\chi)C_{\ell_4}^{T\phi}C_L^{\phi T}(\chi)+C_{\ell_2}^{T\phi}(\chi)C_{\ell_4}^{TT}C_L^{\phi T}(\chi)\right)\bigg],
\eeq
Secondly, the second-order contributions to \eqref{eq: lensing-full-sky-harmonic} induce a contact trispectrum, which we denote ISW-IV:
\beq\label{eq: ISW-contact-trispectra}
    \av{a_{\ell_1m_1}a_{\ell_2m_2}a_{\ell_3m_3}a_{\ell_4m_4}}_c &\supset& \frac{1}{2}\mathcal{J}^{\ell_1m_1}_{\ell_2(-m_2)\ell_3(-m_3)\ell_4(-m_4)}C_{\ell_2}^{TT}C_{\ell_3}^{T\phi}C_{\ell_4}^{T\phi}+\text{23 perms.}
\eeq
Whilst this can also be expressed in reduced form, the resulting expression is lengthy and uninteresting. Each of the four ISW terms peak in a different kinematic limit (as sketched in Fig.\,\ref{fig: cartoon}), with the ISW-I and ISW-IV contributions almost canceling in the doubly squeezed regime of $\ell_1,\ell_2\gg \ell_3,\ell_4$. Generalized trispectra including polarization are presented in Appendix \ref{app: pol}.

As discussed in \S\ref{sec: flat-sky}, the exchange trispectra can be equivalently derived by treating each pair of lensed fields as a quadratic estimator for $\phi_{LM}$ or $a_{LM}$. This involves the partially-contracted two-point functions:
\beq\label{eq: quad-lensing+isw-full}
    \av{a_{\ell_1m_1}a_{\ell_3m_3}}_{\rm unlensed} &=&\sum_{LM}(-1)^{M} \mathcal{I}^{\ell_1m_1}_{LM\ell_3(-m_3)}C_{\ell_3}^{TT}\phi_{LM}+(1\leftrightarrow3)\\\nonumber
    \av{a_{\ell_1m_1}a_{\ell_3m_3}}_{\rm ISW} &=& \sum_{LM}(-1)^{M}\mathcal{I}^{\ell_1m_1}_{LM\ell_3(-m_3)}\int_0^{\chi_\star}d\chi\,C_{\ell_3}^{T\phi}(\chi)a_{LM}(\chi)+(1\leftrightarrow3)
\eeq
analogous to \eqref{eq: quad-lensing}\,\&\,\eqref{eq: quad-ISW}. The auto- and cross-spectra of these immediately leads to the four-point functions of \eqref{eq: tris-ex-full-sky-T}. 

\section{Estimators}\label{sec: estimators}
\noindent How can we constrain the above non-Gaussian signatures using observational data? For lensing, one conventionally builds `quadratic estimators' for the lensing potential exploiting \eqref{eq: quad-lensing+isw-full}, before computing the power spectrum of $\widehat{\phi}_{LM}$ \citep[e.g.,][]{Hu:2001kj,Okamoto:2003zw}. The similarity between the first and second terms in \eqref{eq: tris-ex-full-sky-T} implies that the ISW-I contribution can be extracted in a similar manner, replacing $C_L^{\phi\phi}\to C_L^{TT}$ and $C_\ell^{TT}\to C_{\ell}^{T\phi}$ in the usual lensing estimator. Moreover, the ISW-II and ISW-III contributions can be probed by cross-correlating the resulting measurements of $\phi_{LM}$ and $a_{LM}$; however, it is less clear how to extract the ISW-IV contributions, which involve a contact trispectrum.

Here, we instead adopt the `optimal template estimator' formalism, whereupon one builds a minimum-variance estimator for some scalar amplitude, $A$, by maximizing the perturbative likelihood of the data \citep[e.g.,][]{1997PhRvD..55.5895T,Sekiguchi:2013hza,2011MNRAS.417....2S,2015arXiv150200635S,Hamilton:2005ma,Komatsu:2003iq}. As shown explicitly in \citep{Philcox4pt1}, the associated lensing estimator is equivalent to the usual form, but naturally includes methodological enhancements such as realization-dependent noise subtraction \citep{Namikawa:2012pe}, optimal combination of temperature and polarization \citep{Maniyar:2021msb} and a full experiment-dependent normalization. Furthermore, this allows straightforward computation of the (non-exchange) ISW-IV and bispectrum contributions, extensions to polarization, as well as assessment of primordial non-Gaussianity biases.

In the notation of \citep{Philcox4pt1} (building on \citep{2021PhRvD.103j3504P,Philcox:2021ukg,Philcox:2024rqr,Philcox:2023uwe,Philcox:2023psd,2011MNRAS.417....2S,2015arXiv150200635S,Komatsu:2003iq}), a general estimator for a set of parameters, $\{A_\alpha\}$, appearing only in the $n$-point function of the data, $d$, is given by
\beq\label{eq: gen-estimator}
    \widehat{A}_\alpha[d] &=& \sum_{\beta}\F^{-1}_{\alpha\beta}\widehat{N}_\beta[d]\\\nonumber
    \widehat{N}_\alpha[d] &=& \frac{1}{n!}\sum_{\ell_1\cdots\ell_n m_1\cdots m_n}\frac{\partial\av{a_{\ell_1m_1}\cdots a_{\ell_nm_n}}_c}{\partial A_\alpha}\left(\mathcal{H}^{(n)}_{\ell_1m_1\cdots\ell_nm_n}[\Si d]\right)^*\\\nonumber
    \F_{\alpha\beta} &=& \frac{1}{n!}\sum_{\ell_1\cdots\ell_n m_1\cdots m_n}\left[\left(\frac{\partial\av{a_{\ell_1m_1}\cdots a_{\ell_nm_n}}_c}{\partial A_\alpha}\right)^*\left([\Si \mathsf{P}]_{\ell_1m_1,\ell_1'm_1'}\cdots [\Si \mathsf{P}]_{\ell_nm_n,\ell_n'm_n'}\right)\left(\frac{\partial\av{a_{\ell'_1m'_1}\cdots a_{\ell'_nm'_n}}_c}{\partial A_\beta}\right)\right]^*\\\nonumber
\eeq
where $\widehat{N}$ and $\F$ are a numerator vector and a normalization matrix respectively. Here, we have defined the pointing matrix $\mathsf{P}$, such that $d = \mathsf{P}a+\text{noise}$ for underlying CMB field $a$, as well as some linear weighting scheme $\Si$ and the $n$-th order Hermite tensor $\mathcal{H}^{(n)}$ (e.g., $\mathcal{H}^{(2)}_{\ell_1m_1\ell_2m_2}[x]=x_{\ell_1m_1}x_{\ell_2m_2}-\av{x_{\ell_1m_1}x_{\ell_2m_2}}$). As shown in \citep{2011MNRAS.417....2S,2015arXiv150200635S} (using methods developed for stochastic trace estimation \citep{girard89,hutchinson90}), the normalization can be computed efficiently using Monte Carlo methods given the derivatives
\beq\label{eq: Q-deriv}
    Q_{\ell m,\alpha}[x^{(2)},\cdots,x^{(n)}] &=& \sum_{\ell_2\cdots \ell_nm_2\cdots m_n}\frac{\partial\av{a_{\ell m}a_{\ell_2m_2}\cdots a_{\ell_nm_n}}_c}{\partial A_\alpha}\left(x^{(2)}_{\ell_2m_2}\cdots x^{(n)}_{\ell_nm_n}\right)^*,
\eeq
which satisfy $\sum_{\ell m}x^{(1)*}_{\ell m}Q_{\ell m,\alpha}[x^{(2)},\cdots,x^{(n)}] = n!\,\widehat{\mathcal{N}}_\alpha[x^{(1)},\cdots x^{(n)}]$. An analogous form can be derived including polarization: this simply adds field indices, e.g.,\ $a_{\ell_i m_i}\to a_{\ell_i m_i}^{X_i}$ and sums over all $X_i\in\{T,E,B\}$. 

Estimator \eqref{eq: gen-estimator} has a number of useful properties \citep[e.g.,][]{Philcox4pt1}: (1) it returns zero for Gaussian datasets, provided that the simulations used to estimate the disconnected terms (e.g., those used to compute $\av{h_{\ell_1m_1}h_{\ell_2m_2}}$) have the same covariance as the data; (2) it is unbiased for $A_\alpha\neq 0$ provided that $\{A_\alpha\}$ completely describe the correlator of interest and that the pointing matrix $\mathsf{P}$ is precisely known; (3) it is minimum-variance in the limit of $\Si \to \mathsf{P}^\dagger \Ci$ and Gaussian statistics, where $\Ci$ is the inverse covariance of the data. Of course, property (3) is never quite satisfied in lensing studies (since the fiducial model is non-Gaussian), but is usually a good approximation for \textit{Planck}-like experiments \citep{Philcox4pt2,Philcox4pt3}.

In the optimal limit, the normalization matrix, $\F_{\alpha\beta}$, is equal to the Fisher matrix of the dataset, such that
\beq\label{eq: cramer-rao}
    \mathrm{cov}(\widehat{A}_\alpha,\widehat{A}_\beta) \geq \F^{-1}_{\alpha\beta}.
\eeq
This facilitates efficient forecasting in the presence of non-trivial beams, masks, and beyond. Moreover, the off-diagonal elements of $\F$ encode the bias on some template $\alpha$ induced by an additional component $\gamma$ with non-zero amplitude:
\beq\label{eq: template-biases}
    \Delta A_\alpha =  (\F_{\alpha\gamma}/\F_{\alpha\alpha})A^{\rm fid}_{\gamma}.
\eeq
Importantly, this holds regardless of whether the estimator is optimal. Denoting $\rho_{\alpha\gamma} \equiv \F_{\alpha\gamma}/\sqrt{\F_{\alpha\alpha}\F_{\gamma\gamma}}$ as the cosine between templates $\alpha$ and $\gamma$, this can be rewritten as
\beq\label{eq: template-biases-snr}
    \Delta A_\alpha/\sigma^{\rm ideal}(A_\alpha) =  \rho_{\alpha\gamma}\,(A^{\rm fid}_{\gamma}/\sigma^{\rm ideal}(A_{\gamma}))
\eeq
where the term in parentheses is the idealized signal-to-noise ratio on $A_\gamma$. In \S\ref{sec: current-data}\,\&\,\ref{sec: forecasts}, we use this to assess biases on primordial non-Gaussianity amplitudes from late-time effects.

In the sections below, we will utilize the above forms to define efficient estimators for the lensing- and ISW-lensing contributions, introducing a set of amplitudes, e.g., $A_{\rm lens}$, with fiducial values of unity. Throughout we will remain agnostic to the choice of weighting scheme $\Si$ and pointing matrix $\mathsf{P}$ (which are, in general, experiment specific), and restrict to temperature perturbations, with extension fo polarization presented in Appendix \ref{app: pol}.

\subsection{Bispectrum Estimator}\label{subsec: bspec-estimator}
\noindent The general bispectrum estimator is a special case of \eqref{eq: gen-estimator} \citep[cf.][]{2011MNRAS.417....2S}. Introducing a suite of random fields $\{\delta\}$ with the same covariance as the data, the numerator can be written
\beq\label{eq: b-numerator}
    \widehat{N}_\alpha[d] &=& \widehat{\mathcal{N}}_\alpha[d,d,d] - \left(\av{\widehat{\mathcal{N}}_\alpha[d,\delta,\delta]}_\delta+\text{2 perms.}\right)\\\nonumber
    \widehat{\mathcal{N}}_\alpha[\alpha,\beta,\gamma] &=& \frac{1}{3!}\sum_{\ell_1\ell_2\ell_3 m_1m_2m_3}\frac{\partial\av{a_{\ell_1m_1}a_{\ell_2m_2}a_{\ell_3m_3}}_c}{\partial A_\alpha}[\Si\alpha]^*_{\ell_1m_1}[\Si\beta]^*_{\ell_2m_2}[\Si\gamma]^*_{\ell_3m_3},
\eeq
where the averages can be computed using $N_{\rm disc}\gg 1$ simulations. Inserting the ISW-lensing bispectrum \eqref{eq: bis-full-sky-T} and a scaling amplitude $A^{(3)}_{\rm ISW}$ (equal to $\widehat{A}^{\phi T}$ of \citep{Carron:2022eum}, $\widehat{A}^{T\phi}$ of \citep{Carron:2022eyg,Planck:2013mth,Planck:2015mym}) with unit fiducial amplitude, we find
\beq
    \widehat{\mathcal{N}}_{A^{(3)}_{\rm ISW}}[\alpha,\beta,\gamma] &=& \frac{1}{3!}\sum_{\ell_1\ell_2\ell_3 m_1m_2m_3}\mathcal{I}^{\ell_1m_1}_{\ell_2(-m_2)\ell_3(-m_3)}C^{T\phi}_{\ell_2}C^{TT}_{\ell_3}[\Si\alpha]^*_{\ell_1m_1}[\Si\beta]^*_{\ell_2m_2}[\Si\gamma]^*_{\ell_3m_3}+\text{5 perms.},
\eeq
which is simply a cubic combination of the data (and simulations) weighted by a template-specific factor. Inserting \eqref{eq: quadratic-kernel} and simplifying, we obtain
\beq\label{eq: bis-ISW-num}
    \widehat{\mathcal{N}}_{A^{(3)}_{\rm ISW}}[\alpha,\beta,\gamma] &=& -\frac{1}{12}\sum_{\lambda=\pm1}\int d\hn\,U[\Si \alpha](\hn)V^{\rm ISW}_{-\lambda}[\Si \beta](\hn)V^{\rm lens}_{\lambda}[\Si \gamma](\hn)+\text{5 perms.}
\eeq
defining the spin-0 and spin-$(-\lambda)$ maps
\beq\label{eq: UV-maps-T}
    U[x](\hn) \equiv&& \sum_{\ell m}Y_{\ell m}(\hn)x_{\ell m}\\\nonumber
    V^{\rm lens}_\lambda[x](\hn) \equiv \sum_{\ell m}{}_{-\lambda}Y_{\ell m}(\hn)\sqrt{\ell(\ell+1)}C_\ell^{TT}x_{\ell m},&& \qquad V^{\rm ISW}_\lambda[x](\hn,\chi) \equiv \sum_{\ell m}{}_{-\lambda}Y_{\ell m}(\hn)\sqrt{\ell(\ell+1)}C_\ell^{T\phi}(\chi)x_{\ell m},
\eeq
where we will only need $V_\lambda^{\rm ISW}[x](\hn,\chi_\star)\equiv V_\lambda^{\rm ISW}[x](\hn)$ in this section. $U,V_\lambda^{\rm lens}$ match the definitions of \citep{Philcox4pt1}, with $U^* = U$, $V^*_\lambda = -V_{-\lambda}$. Notably, \eqref{eq: bis-ISW-num} is explicitly separable and can be computed with one spin-zero spherical harmonic transform, two pairs of spin-one transforms and a sum over pixels. This can be rewritten using the unnormalized lensing estimator $\Phi_{LM}$ defined in \S\ref{subsec: tspec-estimator}:
\beq
    \widehat{\mathcal{N}}_{A^{(3)}_{\rm ISW}}[\alpha,\beta,\gamma] &=& \frac{1}{6}\sum_{LM}\sqrt{L(L+1)}C_L^{T\phi}[\Si \beta]^*_{LM}\Phi^{\rm lens}_{LM}[\Si \alpha,\Si\gamma]+\text{5 perms.};
\eeq
this is just the cross-spectrum of a temperature map with the reconstructed lensing potential, demonstrating equivalence with previous estimators \citep{Carron:2022eum,Lewis:2011fk,Mangilli:2013sxa,Carron:2022eyg}.

As discussed in \citep{2011MNRAS.417....2S,Philcox:2021ukg,Philcox:2024rqr,Philcox:2023uwe}, the Fisher matrix can be efficiently computed using Monte Carlo methods. Introducing a set of $N_{\rm fish}$ (Gaussian) random maps $\{a^{(i)}\}$ with $i\in\{1,2\}$ and invertible covariance $\mathsf{A}$, we can write
\beq\label{eq: fish3}
    \mathcal{F}_{\alpha\beta} &=& \frac{1}{4}\left(F^{11,11}_{\alpha\beta}+F^{22,22}_{\alpha\beta}-F^{11,22}_{\alpha\beta}-F^{22,11}_{\alpha\beta}\right)\\\nonumber
    F^{ab,cd}_{\alpha\beta} &\equiv& \frac{1}{6}\sum_{\ell m\ell'm'}\bigg\langle\left(Q^*_{\ell m,\alpha}[\Si \mathsf{P}a^{(a)},\Si\mathsf{P} a^{(b)}]\right)[\mathsf{S}^{-1}\mathsf{P}]_{\ell m,\ell'm'}\left(Q_{\ell'm',\beta}[\mathsf{A}^{-1} a^{(c)},\mathsf{A}^{-1} a^{(d)}]\right)\bigg\rangle^*_a,
\eeq
which involves the $Q$ derivatives of \eqref{eq: Q-deriv}.\footnote{By taking expectations the random fields, with $\av{\Ai aa^\dagger}_a = \mathsf{I}$, it is straightforward to show that this reproduces \eqref{eq: gen-estimator}.} For the ISW-lensing bispectrum, these are given by
\beq\label{eq: Q-deriv-bis}
    Q_{\ell m,A^{(3)}_{\rm ISW}}[x,y] &=& \frac{1}{2}\sum_{\lambda=\pm1}\sqrt{\ell(\ell+1)}\int d\hn\,{}_{\lambda}Y^*_{\ell m}(\hn)U[x](\hn)\left(C_{\ell}^{T\phi}V^{\rm lens}_{-\lambda}[y](\hn)+C_\ell^{TT}V^{\rm ISW}_{-\lambda}[x](\hn)\right)\\\nonumber
    &&\,-\,\frac{1}{2}\sum_{\lambda=\pm1}\int d\hn\,Y^*_{\ell m}(\hn)V^{\rm ISW}_{-\lambda}[x]V^{\rm lens}_\lambda[y](\hn)\,+\,(x\leftrightarrow y).
\eeq
As for the numerator, this can be computed using one spin-zero and two pairs of spin-one harmonic transforms, with the full Fisher matrix computed as an $\ell$-space sum (after applying the pointing matrix, which usually involves further transforms).\footnote{Assuming translation-invariant noise and unit mask, the Fisher matrix can also be computed analytically. This is obtained by setting $\F_{\alpha\beta} = \mathrm{cov}(\widehat{N}_\alpha,\widehat{N}_\beta)$ (true under idealized assumptions) and using the relation
\beq
    \sum_{mm'} {}_{s}Y_{\ell m}(\hn){}_{s'}Y^*_{\ell'm'}(\hn')\av{x_{\ell m}^*y_{\ell'm'}} &=& \delta^{\rm K}_{\ell\ell'}C_{\ell}^{xy}(-1)^{s}\frac{2\ell+1}{4\pi}d^\ell_{ss'}(\theta)
\eeq
for $\theta = \cos^{-1}(\hn\cdot\hn')$ and Wigner $d$ symbols $d_{ss'}^{\ell}$ \citep{2015arXiv150200635S}. This simplifies the $\hn,\hn'$ integrals into a one-dimensional integral over $\cos\theta$.} Computation of both the numerator and normalization thus scales as $\mathcal{O}(\ell_{\rm max}^2\log\ell_{\rm max})$, with an additional $N_{\rm disc}$ ($N_{\rm fish}$) factor for the numerator (normalization), due to the Monte Carlo summation.

\subsection{Trispectrum Estimator}\label{subsec: tspec-estimator}
\noindent Next, we present the lensing and ISW-lensing trispectrum estimators (with the former matching \citep{Philcox4pt1}). To connect with previous estimators and to build intuition, we first define quadratic estimators for lensing and temperature fields. These can be obtained by projecting the (pre-symmetrized derivatives of the) unlensed-averaged and ISW-averaged correlators of \eqref{eq: quad-lensing+isw-full} onto two copies of the filtered data:
\beq\label{eq: lens-isw-quad-estimators}
    \Phi^{\rm lens}_{LM}[x,y] &\equiv& \frac{1}{\sqrt{L(L+1)}}\sum_{\ell_1\ell_3m_1m_3}\mathcal{I}^{\ell_1m_1}_{L(-M)\ell_3(-m_3)}C_{\ell_3}^{TT}x^*_{\ell_1m_1}y^{*}_{\ell_3m_3} =  \frac{1}{2}\sum_{\lambda=\pm1}\int d\hn\,{}_{\lambda}Y^*_{LM}(\hn)U[x](\hn)V_{-\lambda}^{\rm lens}[y]\\\nonumber
    \Phi^{\rm ISW}_{LM}[x,y](\chi) &\equiv& \frac{1}{\sqrt{L(L+1)}}\sum_{\ell_1\ell_3m_1m_3}\mathcal{I}^{\ell_1m_1}_{L(-M)\ell_3(-m_3)}C_{\ell_3}^{T\phi}(\chi)x^*_{\ell_1m_1}y^{*}_{\ell_3m_3} =  \frac{1}{2}\sum_{\lambda=\pm1}\int d\hn\,{}_{\lambda}Y^*_{LM}(\hn)U[x](\hn)V_{-\lambda}^{\rm ISW}[y](\chi),
\eeq
where $U$, $V^{\rm ISW}$ and $V^{\rm lens}$ were defined in \eqref{eq: UV-maps-T}. At leading-order, these are biased estimators for $\phi_{LM}$ and $a_{LM}(\chi)$ respectively, and can be easily implemented using spin-one transforms. Due to the similarity of the underlying lensing effect \eqref{eq: lensing-full-sky-harmonic}, the two estimators differ only by the weighting: $C_{\ell}^{TT}$ or $C_{\ell}^{T\phi}(\chi)$. 

The full exchange trispectrum estimators can be derived by squaring \eqref{eq: lens-isw-quad-estimators} and adding various bias and normalization terms \citep[e.g.,][]{Carron:2022edh}. Equivalently, we can insert the theoretical trispectrum of \eqref{eq: tris-ex-full-sky-T} into the general $n$-point function estimator \eqref{eq: gen-estimator}, whose numerator takes the following form for $n=4$:
\beq\label{eq: estimator-n=4}
    \widehat{N}_\alpha[d] &=& \widehat{\mathcal{N}}_{\alpha}[d,d,d,d] - \left(\av{\widehat{\mathcal{N}}_{\alpha}[d,d,\delta,\delta]}_\delta+\text{5 perms.}\right)+\left(\av{\widehat{\mathcal{N}}_{\alpha}[\delta^{(1)},\delta^{(1)},\delta^{(2)},\delta^{(2)}]}_{\delta^{(1)},\delta^{(2)}}+\text{2 perms.}\right)\nonumber\\
    \widehat{\mathcal{N}}_\alpha[\alpha,\beta,\gamma,\delta] &=& \frac{1}{4!}\sum_{\ell_1\cdots\ell_4 m_1\cdots m_4}\frac{\partial\av{a_{\ell_1m_1}\cdots a_{\ell_4m_4}}_c}{\partial A_\alpha}[\Si \alpha]^*_{\ell_1m_1}\cdots [\Si \delta]^*_{\ell_4m_4}.
\eeq
Here, we have introduced two uncorrelated sets of simulations, $\{\delta^{(1)}\}$ and $\{\delta^{(2)}\}$, whose covariance match the data -- these both remove the mean-field of the estimator (\textit{i.e.}\ $\av{\Phi}\neq 0$) and perform realization-dependent bias subtraction \citep[e.g.,][]{Namikawa:2012pe}. Introducing a lensing amplitude $A_{\rm lens}$ with a fiducial value of unity, we find the lensing numerator: 
\beq\label{eq: lens-tspec-estimator-T}
    \widehat{\mathcal{N}}_{A_{\rm lens}}[\alpha,\beta,\gamma,\delta] = \frac{1}{24}\sum_{LM}(-1)^{M}L(L+1)C_L^{\phi\phi}\Phi^{\rm lens}_{LM}[\Si\alpha,\Si\beta]\Phi^{\rm lens}_{L(-M)}[\Si\gamma,\Si\delta]+\text{11 perms.},
\eeq
which matches previous works \citep{Carron:2022edh,Philcox4pt1}.  The estimators for the ISW-I and ISW-II\,+\,ISW-III amplitudes are analogous:
\beq\label{eq: isw-ex-tspec-estimator-T}
    \widehat{\mathcal{N}}_{A_{\rm ISW-I}}[\alpha,\beta,\gamma,\delta] &=& \frac{1}{24}\sum_{LM}(-1)^{M}L(L+1)C_L^{TT}\Phi^{\rm ISW}_{LM}[\Si\alpha,\Si\beta]\Phi^{\rm ISW}_{L(-M)}[\Si\gamma,\Si\delta]+\text{11 perms.}\\\nonumber
    \widehat{\mathcal{N}}_{A_{\rm ISW-II+III}}[\alpha,\beta,\gamma,\delta] &=& \frac{1}{24}\sum_{LM}(-1)^{M}L(L+1)\int_0^{\chi_\star}d\chi\,C_L^{\phi T}(\chi)\bigg(\Phi^{\rm ISW}_{LM}(\chi)[\Si\alpha,\Si\beta]\Phi^{\rm lens}_{L(-M)}[\Si\gamma,\Si\delta]\\\nonumber
    &&\qquad\qquad\qquad\qquad\qquad\qquad\,+\,\Phi^{\rm lens}_{LM}[\Si\alpha,\Si\beta]\Phi^{\rm ISW}_{L(-M)}[\Si\gamma,\Si\delta](\chi)\bigg)+\text{11 perms.},
\eeq
where $A_{\rm ISW-I}$ and $A_{\rm ISW-II+III}$ encode the ratio of $C^{T\phi}_{\ell_3}C_{\ell_4}^{T\phi}C_L^{TT}$ and $\int d\chi\,(C^{T\phi}_{\ell_3}(\chi)C_{\ell_4}^{TT}+C^{TT}_{\ell_3}C_{\ell_4}^{T\phi}(\chi))C_L^{\phi T}(\chi)$, to their fiducial value, and we note that the approximation $\chi\approx \chi_\star$ is not valid for ISW-II and ISW-III. As expected, these estimators are simply the power spectra of \eqref{eq: lens-isw-quad-estimators}, weighted by the theoretical expectations, and can be straightforwardly computed via summation in harmonic-space.

In practice, the estimators for ISW-II and ISW-III are more expensive to implement than for ISW-I, due to the explicit $\chi$ integral. Noting that these terms are generally expected to be small (since they arise from lensing of the ISW, rather than lensing of the primary CMB) and foreshadowing the conclusions of \S\ref{sec: current-data}, we here adopt a simple approximation motivated by numerical computation of the full trispectrum. In particular, we assume
\beq\label{eq: chi-approx}
    \int_0^{\chi_\star} d\chi\,C_\ell^{T\phi}(\chi)C_L^{\phi T}(\chi)\approx \int_0^{\chi_\star} d\chi\,C_\ell^{T\phi}(\chi_{\rm eff})C_L^{\phi T}(\chi) \equiv C_\ell^{T\phi}(\chi_{\rm eff})C_L^{T\phi},
\eeq
where $\chi_{\rm eff}$ corresponds to the distance at which the ISW source is maximized. Here, we fix $z(\chi_{\rm eff})= 0.8$, which minimizes the error in \eqref{eq: chi-approx} for small $\ell,L$.\footnote{An alternative approach would be to assume $\chi_{\rm eff}=\chi_\star$; due to the shape of the lensing kernel, this results in a strict overestimation of the ISW-II and ISW-III contributions (by a factor $\approx 3$).} This significantly expedites computation without changing our overall conclusions.

Due to its cubic source, the numerator of the ISW-IV estimator takes a somewhat different form. Inserting \eqref{eq: tris-con-full-sky-T} into \eqref{eq: estimator-n=4} and simplifying, we find
\beq\label{eq: isw-con-tspec-estimator-T}
    \widehat{\mathcal{N}}_{A_{\rm ISW-IV}}[\alpha,\beta,\gamma,\delta] &=&\frac{1}{96}\sum_{\lambda=\pm1}\int d\hn\,U[\Si\alpha]V^{\rm ISW}_\lambda[\Si\delta](\hn)\\\nonumber
    &&\,\times\,\bigg[V^{\rm ISW}_{-\lambda}[\Si\gamma](\hn)S^{\rm lens}_0[\Si\beta](\hn)+V^{\rm ISW}_\lambda[\Si\gamma](\hn)S^{\rm lens}_\lambda[\Si\beta](\hn)\bigg]+\text{11 perms.},
\eeq
where $A_{\rm ISW-IV}$ is proportional to $C_{\ell_2}^{TT}C_{\ell_3}^{T\phi}C_{\ell_4}^{T\phi}$. This involves the spin-$2\lambda$ field 
\beq\label{eq: S-lens-def}
    S^{\rm lens}_\lambda[x](\hn)&\equiv&\sum_{\ell m}C^{TT}_{\ell}\sqrt{\ell(\ell+1)(\ell-|\lambda|)(\ell+|\lambda|+1)}{}_{2\lambda}Y_{\ell m}(\hn)x_{\ell m},
\eeq 
with $\left(S^{\rm lens}_{\lambda}\right)^* = S^{\rm lens}_{-\lambda}$. This is analogous to the bispectrum estimator, and can be computed as a pixel-space summation following spin-zero, spin-one and spin-two harmonic transforms.

Finally, we require the normalization of the estimators, \textit{i.e.}\ the Fisher matrix $\F$. Following \citep{2015arXiv150200635S,Philcox4pt1}, this can be computed using Monte Carlo tricks analogous to \eqref{eq: fish3} for the three-point function; here, this requires
\beq\label{eq: fish4}
    \F_{\alpha\beta} &=& \frac{1}{48}\bigg[\left(F^{111,111}_{\alpha\beta}+F^{222,222}_{\alpha\beta}\right)\,+\,9\left(F^{112,112}_{\alpha\beta}+F^{122,122}_{\alpha\beta}\right)\,-\,3\left(F^{111,122}_{\alpha\beta}+F^{222,112}_{\alpha\beta}+F^{122,111}_{\alpha\beta}+F^{112,222}_{\alpha\beta}\right)\bigg]\nonumber\\
    F^{abc,def}_{\alpha\beta} &\equiv& \frac{1}{24}\sum_{\ell m\ell'm'}\bigg\langle\left(Q^*_{\ell m,\alpha}[\Si \mathsf{P}a^{(a)},\Si\mathsf{P} a^{(b)},\Si\mathsf{P} a^{(c)}]\right)[\mathsf{S}^{-1}\mathsf{P}]_{\ell m,\ell'm'}\nonumber\\
    &&\qquad\qquad\qquad\,\times\,\left(Q_{\ell'm',\beta}[\mathsf{A}^{-1} a^{(d)},\mathsf{A}^{-1} a^{(e)},\mathsf{A}^{-1}a^{(f)}]\right)\bigg\rangle^*_a,
\eeq
where $\{a^{(1)}\},\{a^{(2)}\}$ are two independent sets of mean-zero random fields with known covariance $\mathsf{A}$. This can be computed as a harmonic-space product, averaging over $N_{\rm fish}$ Monte Carlo realizations, given the relevant $Q_{\ell m,\alpha}$ functions. In practice, these can be computed by inserting the relevant trispectrum definitions into \eqref{eq: Q-deriv} and simplifying, which yields similar forms to those found for the bispectrum \eqref{eq: Q-deriv-bis}. Since the full expressions are lengthy and generally uninformative, they are relegated to Appendix \ref{app: Q-derivs}; here, we note that each expression can be implemented entirely using spin-weighted spherical harmonic-transforms and summation in pixel- or harmonic-space. This yields an estimator with complexity $\mathcal{O}(\ell_{\rm max}^2\log\ell_{\rm max})$ as for the three-point function.

\section{Application to Current Data}\label{sec: current-data}
\noindent Given the bispectrum and trispectrum estimators presented in \S\ref{sec: estimators}, we can now search for ISW-lensing correlators in observational data. Here, we analyze the latest data from \textit{Planck} (Public Release 4; PR4), processed with the \textsc{npipe} pipeline, alongside a suite of $200$ FFP10/\textsc{npipe} simulations. All analysis choices are identical to that of \citep{Philcox4pt3}, including the mask (with $f_{\rm sky}=0.68$), beam, idealized $\Si$ weighting scheme,\footnote{We additionally test the optimal weighting scheme, $\Si_{\rm opt}$. This tightens constraints by $5-10\%$ (as in \citep{Mangilli:2013sxa}) but significantly increases computation time due to the conjugate gradient descent algorithm used to approximately invert the covariance matrix.} estimator hyperparameters, and fiducial cosmology. To elucidate any bias from residual foregrounds, we utilize two choices of component-separation pipeline: \textsc{sevem} and \textsc{smica}. By default, we include all modes with $\ell\in[2,2048]$, though assess the dependence on $\ell_{\rm max}$ below. Half of the simulations are used to subtract the Gaussian contributions to the estimator, whilst the remaining set are used to estimate the empirical variances. For the bispectrum estimators, we will include both temperature and polarization information; given our eventual conclusions, we restrict only to temperature-modes in trispectrum studies. 

All of the estimators used below have been implemented within the \href{https://github.com/oliverphilcox/PolySpec}{\textsc{PolySpec}} package described in \citep{Philcox4pt2}. For the temperature-only ISW and $\fnl$ bispectra at the \textit{Planck} resolution, computation of $N_{\rm fish}=20$ Fisher matrix realizations requires $6$ node-minutes, with a further $12$ node-minutes required to process $100$ simulations with $N_{\rm disc}=100$. These values approximately double when including polarization, and increase by a factor of a few when computing trispectra (as discussed in \citep{Philcox4pt2}). 

\begin{table}[t]
    \centering
    \begin{tabular}[t]{c||l|ccc|ccc|cc}
    \multicolumn{10}{c}{\textbf{Bispectrum}}\\
    & Field & \multicolumn{3}{c|}{\textsc{sevem}} & \multicolumn{3}{c|}{\textsc{smica}} & \multicolumn{2}{c}{Fisher}\\\hline
    & $T$ & $0.84$ & $\pm$ & $0.29$ & $0.96$ & $\pm$ & $0.29$ & $\pm$ & $0.24$\\
    $\boldsymbol{A_{\rm ISW}^{(3)}}$ & $E,B$ & $-2.6$ & $\pm$ & $3.5$ & $-1.4$ & $\pm$ & $3.6$ & $\pm$ & $3.2$\\
    & $T,E,B$ & $0.86$ & $\pm$ & $0.23$ & $0.90$ & $\pm$ & $0.22$ & $\pm$ & $0.17$\\
    \end{tabular}
    \hskip15pt
    \begin{tabular}[t]{c||c|ccc|ccc|cc}
    \multicolumn{10}{c}{\textbf{Trispectrum}}\\
    & Field & \multicolumn{3}{c|}{\textsc{sevem}} & \multicolumn{3}{c|}{\textsc{smica}} & \multicolumn{2}{c}{Fisher}\\\hline
    $\boldsymbol{A_{\rm lens}}$ & $T$ & $0.964$ & $\pm$ & $0.028$ & $0.959$ & $\pm$ & $0.028$ & $\pm$ & $0.020$\\\hline
    $\boldsymbol{A^{(4)}_{\rm ISW}}$ & $T$ & $-0.6$ & $\pm$ & $18.6$ & $-3.8$ & $\pm$ & $18.1$ & $\pm$ & $17.1$\end{tabular}
    \caption{\textbf{Left}: Constraints on the ISW-lensing bispectrum amplitude, $A_{\rm ISW}^{(3)}$, obtained from \textit{Planck} PR4 temperature and polarization anisotropies. All results are obtained at $\ell_{\rm max}=2048$, with variances computed from $100$ FFP10 simulations -- these are somewhat weaker than the theoretical errors due to lensing-induced non-Gaussianity. In all cases, we find good agreement with the fiducial value of unity, matching previous works \citep{Planck:2013mth,Planck:2015mym,Carron:2022eyg,Carron:2022eum}. 
    \textbf{Right}: Analogous constraints on the trispectrum amplitudes describing lensing ($A_{\rm lens}$) and ISW-lensing ($A_{\rm ISW}^{(4)}$) from \textit{Planck} PR4 temperature anisotropies. Whilst lensing is detected at high significance (as in previous works \citep[e.g.,][]{Carron:2022eyg,Philcox4pt3}), the data are unable to constrain the ISW-lensing four-point function.}
    \label{tab: bis-tris-results}
\end{table}

\subsection{Bispectra}\label{subsec: results-bspec}

\noindent First, we place constraints on the ISW-lensing bispectrum, parametrized by $A^{(3)}_{\rm ISW}$. As shown in Tab.\,\ref{tab: bis-tris-results}, we find a fairly significant detection of the effect in \textit{Planck} data, with the \textsc{sevem} (\textsc{smica}) temperature-only constraint $2.8\sigma$ ($3.4\sigma$) above zero. Whilst the polarization channels are not sufficiently constraining to detect the fiducial model, their combination with temperature leads to a tight constraint:
\beq
    A_{\rm ISW}^{(3)} = 0.86\pm0.23\quad(\textsc{sevem}) \quad = 0.90\pm0.22 \quad (\textsc{smica}),
\eeq
which is non-zero at $3.8\sigma$ ($4.1\sigma$). We find good agreement between \textsc{sevem} and \textsc{smica} ($<0.5\sigma$, or $<0.2\sigma$ in the combined analysis), indicating that our results are not strongly affected by foreground residuals. Furthermore, these results are in good agreement with previous studies: \textit{Planck} PR2 found $A_{\rm ISW}^{(3)}= 0.90\pm0.28$  ($0.68\pm0.32$) including (excluding) polarization \citep{Planck:2015mym}, which was tightened to $0.94\pm0.30$ and $1.01\pm0.25$ using \textit{Planck} PR3 and PR4 temperature-plus-polarization \textsc{smica} data in \citep{Carron:2022eyg}. Slight differences arise due to the finite number of Monte Carlo simulations, our wider scale ranges (with $\ell_{\min}=8$ in the former works), and our full mask-dependent normalization.

\begin{figure}[!t]
    \centering
    \includegraphics[width=0.6\linewidth]{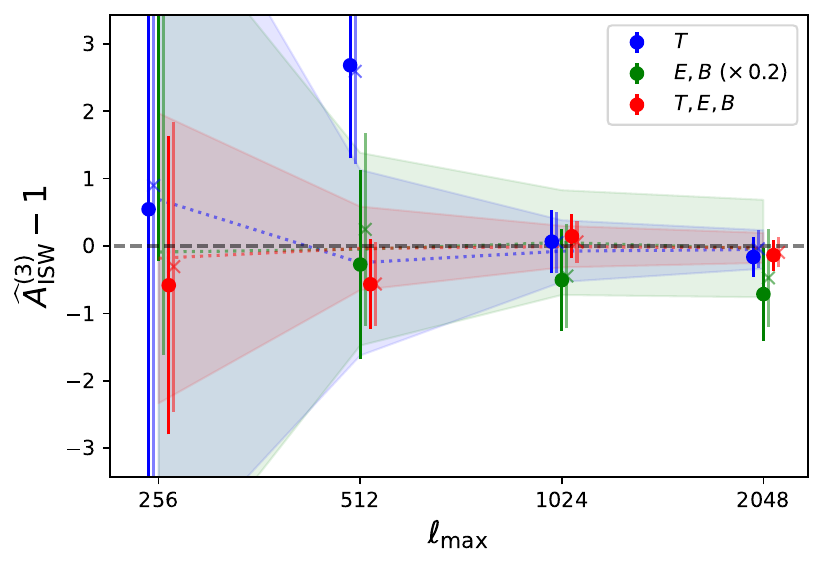}
    \caption{\textit{Planck} constraints on the ISW-lensing bispectrum amplitude as a function of scale. We show results for three choices of field (indicated by color), with dark circles (light crosses) giving \textsc{sevem} (\textsc{smica}) constraints. The solid bands and dotted lines show the mean and errors obtained from $100$ FFP10 mocks. We divide the polarization-only constraints by a factor of five for visibility. All results are consistent with the fiducial amplitude, and the $\ell_{\rm max}=2048$ constraints are summarized in Tab.\,\ref{tab: bis-tris-results}.}
    \label{fig: bis-results}
\end{figure}

In Fig.\,\ref{fig: bis-results}, we plot the constraints as a function of $\ell_{\rm max}$, finding a sharp dependence in all cases. This matches expectation, since the ISW-lensing bispectra are strongly squeezed (\S\ref{sec: flat-sky}), with $\ell_{\rm max}$ setting the precision of the reconstructed lensing field. We find somewhat weaker scalings for the polarization-only analysis, with constraints almost saturating by $\ell_{\rm max}=1024$. This is due to their increased noise relative to the temperature field.

To validate the above results, we first analyze $100$ lensed FFP10 simulations. For all combinations of temperature and polarization we recover the fiducial amplitude to within $0.16\sigma = 1.6\sigma_{\rm mean}$; this implies that we are robust to any contaminants present within the simulations such as residual foregrounds. As seen in Tab.\,\ref{tab: bis-tris-results}, we find slightly ($\lesssim 30\%$) larger variances in the simulations compared to the Fisher matrix predictions -- this could arise either from finite-mock effects or additional sources of covariance, such as lensing non-Gaussianity. Computing the normalization using $N_{\rm fish}=10$ instead of $N_{\rm fish}=20$ leads to negligible change in our constraints, with a mean absolute shift in $\widehat{A}_{\rm ISW}^{(3)}$ of $0.02\sigma$ and a $<0.5\%$ change in the errorbar. Changing the number of Monte Carlo simulations in the numerator has a larger effect \citep[cf.][]{Philcox4pt2}, with $0.1\sigma$ change in the \textit{Planck} constraint obtained using $N_{\rm disc}=50$ (similar across all simulations), but $<0.02\sigma$ for $N_{\rm disc}=200$, implying that our fiducial choice is robust.

\subsection{Trispectra}
\begin{figure}
    \centering
    \includegraphics[width=0.45\linewidth]{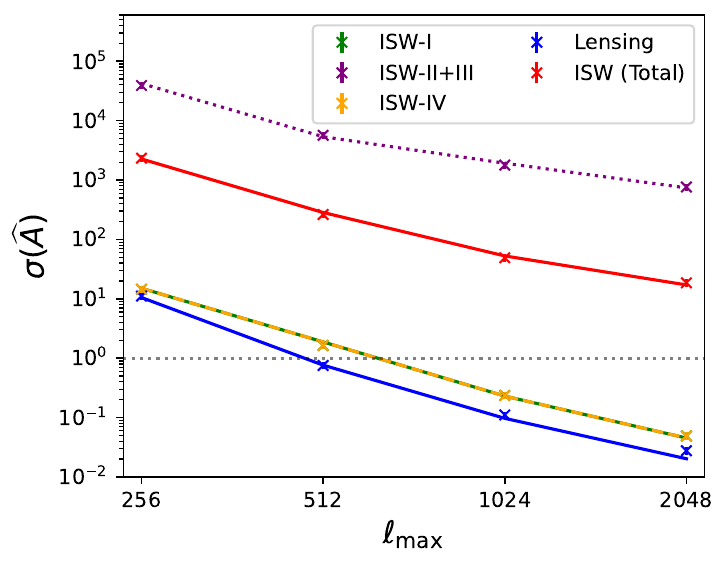}
    \includegraphics[width=0.54\linewidth]{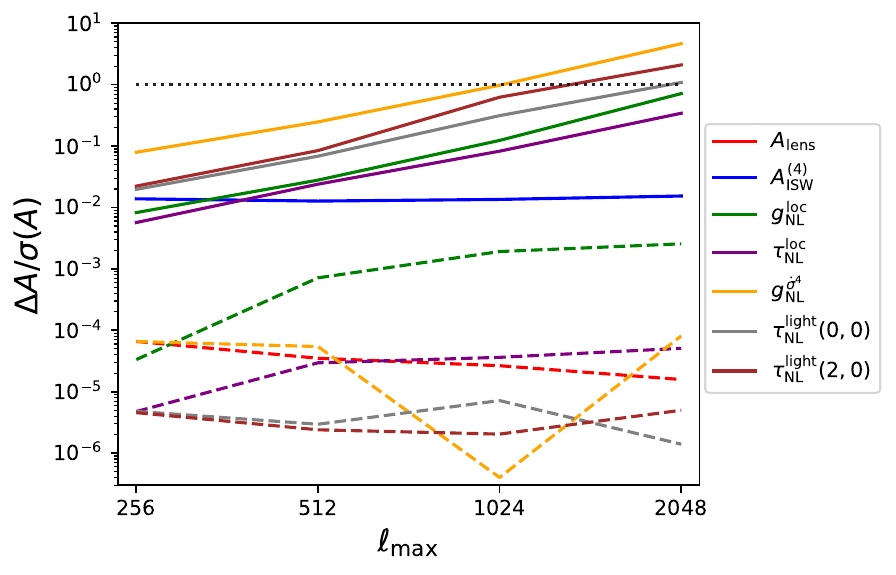}
    \caption{\textbf{Left}: \textit{Planck} PR4 constraints on the lensing and ISW-lensing trispectrum amplitudes as a function of scale, $\ell_{\rm max}$. We split the ISW-lensing trispectrum into the ISW-I, ISW-II/ISW-III and ISW-IV contributions as in \S\ref{sec: flat-sky}, with ISW-II and ISW-III sourced by lensing of the ISW effect itself. Crosses indicate the errorbars obtained from 100 \textsc{sevem} FFP10 simulations, whilst lines give the theoretical variances. Whilst the ISW-I and ISW-IV contributions (green and yellow) are individually large, the combination (red) is highly suppressed, leading to the non-detection in Tab.\,\ref{tab: bis-tris-results}. \textbf{Right}: Fractional biases on primordial and late-time trispectrum amplitudes induced by lensing (solid lines) and ISW-lensing (dashed lines) as a function of scale. All biases are computed for a \textit{Planck}-like survey using \eqref{eq: template-biases}, including the observational mask, beam and noise. 
    The ISW-lensing effect leads to negligible biases on all templates.}
    \label{fig: tris-results}
\end{figure}

\noindent Constraints on the lensing and ISW-lensing four-point functions may be derived similarly. Our key results are shown in Tab.\,\ref{tab: bis-tris-results} \& Fig.\,\ref{fig: tris-results}, restricting to temperature-modes in all cases. As in previous works (including \citep{Carron:2022eyg,Philcox4pt3} for \textit{Planck} PR4), we obtain a strong detection of $A_{\rm lens}$:
\beq
    A_{\rm lens} = 0.964\pm0.028 \quad (\textsc{sevem}) \quad = 0.959\pm0.028 \quad (\textsc{smica}),
\eeq
reaching $34\sigma$, and compatible with the fiducial amplitude within $1.5\sigma$. In contrast, we find only weak bounds on the combined ISW-lensing amplitude (including all four contributions shown in Fig.\,\ref{fig: cartoon}):
\beq
    A_{\rm ISW}^{(4)} = -0.6\pm18.6 \quad (\textsc{sevem}) \quad = -3.8\pm18.1 \quad (\textsc{smica}).
\eeq
This paints a somewhat depressing picture -- to obtain a $3\sigma$ detection of $A_{\rm ISW}^{(4)}$ we would require data with around $50\times$ higher precision. Whilst the polarization sector can also be used to constrain such effects (as discussed in Appendix \ref{app: pol}), the analogous improvements on the bispectrum amplitude $A_{\rm ISW}^{(3)}$ indicate that the full \textit{Planck} dataset will still be unable to meaningfully bound the ISW-lensing trispectrum, thus we do not attempt a polarization analysis in this work.

Fig.\,\ref{fig: tris-results} shows the errorbar on the trispectrum amplitudes as a function of scale, splitting the ISW-lensing amplitude into the four physical contributions described in \S\ref{sec: flat-sky}. As for the bispectrum, the amplitudes are strongly sensitive to scale-cuts, with the lensing and combined ISW amplitudes displaying $4.8\times$ and $3.1\times$ increases in signal-to-noise from $\ell_{\rm max}=1024$ to $\ell_{\rm max}=2048$, despite the inherent noise limitations of \textit{Planck}. As predicted above, the late-time lensing terms (ISW-II and ISW-III) are very small with a theoretical signal-to-noise-ratio (SNR) of $0.0014$ at $\ell_{\rm max}=2048$. In contrast, the ISW-I and ISW-IV contributions are individually large, 
with a forecasted detection significance of $22\sigma$. Whilst this would na\"ively imply that the ISW-lensing signal should be large enough to observe in \textit{Planck} data, the ISW-I (exchange) and ISW-IV (contact) contributions are almost perfectly degenerate (to within $10^{-3}\%$), thus their sum is negligible, with a maximal SNR of just $0.06$. This matches the analytic arguments of \S\ref{sec: flat-sky}, and underscores the importance of including higher-order terms in the lensing expansion.

Late-time trispectra can induce significant biases on primordial amplitudes and, potentially, spurious detections of non-Gaussianity \citep[e.g.,][]{2015arXiv150200635S}. In the right panel of Fig.\,\ref{fig: tris-results}, we show the late-time biases on a set of primordial trispectrum amplitudes: quadratic and cubic local non-Gaussianity ($\taunl$ and $\gnl$), equilateral single-field non-Gaussianity ($\gnldotdot$), conformally-coupled scalar ($\tau_{\rm NL}^{\rm light}(0,0)$) and spin-two ($\tau_{\rm NL}^{\rm light}(2,0)$) collider non-Gaussianity.\footnote{We restrict to a representative set of models that were shown to correlate significantly with lensing in \citep{Philcox4pt3}. Full descriptions of each model and the corresponding estimators can be found in \citep{Philcox4pt1}.} Whilst the large errorbars imply that all parameters are unbiased at low $\ell_{\rm max}$, lensing induces notable biases for higher-resolution studies which require careful subtraction; these reach $5\sigma,2\sigma,1\sigma$ for $\gnldotdot,\tau_{\rm NL}^{\rm light}(2,0),\tau_{\rm NL}^{\rm light}(0,0)$ at the full \textit{Planck} resolution. In contrast, ISW-lensing does not lead to noticeable biases in any parameters, with a maximum shift of $0.003\sigma$ in $\gnl$. This is a consequence of the low signal-to-noise ratio on $A_{\rm ISW}^{(4)}$, following \eqref{eq: template-biases-snr}.

Finally, we perform a number of consistency tests to check that our analysis is robust, as in \S\ref{subsec: results-bspec}. Applying the pipeline to $100$ FFP10 simulations, we find results consistent with the fiducial amplitudes to within $0.4\sigma$, though we find a $30\%$ broader error on $A_{\rm lens}$ than the Fisher forecast (as seen in the left panel of Fig.\,\ref{fig: bis-results}), due to likelihood non-Gaussianity. Halving the number of Monte Carlo simulations in the normalization changes results by $0.1\sigma$ (for lensing) or $0.01\sigma$ (for ISW-lensing), with $\approx 0.1\sigma$ shifts seen when varying the number of numerator simulations. These values imply that our fiducial results are stable to modeling choices and thus our conclusions are robust.

\section{Future Forecasts}\label{sec: forecasts}
\noindent 
Finally, we forecast the detectability of the ISW-lensing correlations in an idealized future survey. For simplicity, we will assume a cosmic-variance-limited full-sky survey ($f_{\rm sky}=1$) with a unit beam and mask.\footnote{For stability, we add a small amount of white-noise, with amplitude $\Delta_T = \Delta_P/\sqrt{2} = 10^{-2}\,\mu\mathrm{K}$-arcmin.} Whilst this set-up is not representative of any upcoming experiment, it provides a useful upper-bound on the signals' detectability. As in \eqref{eq: cramer-rao}, the theoretical errorbars can be obtained from the Fisher matrix, which is computed as a summation over $N_{\rm fish}=20$ Monte Carlo iterations.\footnote{Whilst the idealized Fisher matrices could be computed analytically \citep{2011MNRAS.417....2S}, the numerical scheme is significantly faster at large $\ell_{\rm max}$.} This requires a few node-hours at $\ell_{\rm max}=6144$ both for the bispectrum and trispectrum. To assess the degradation in constraining power due to lensing non-Gaussianity, we additionally analyze a set of $50$ FFP10 simulations (with $\ell_{\rm max}$ set by the simulation resolution), which include the beam and mask. For the ISW-II and III trispectrum contributions, we set $\chi_{\rm eff}=\chi_\star$ (cf.\,\ref{eq: chi-approx}), which provides a (slight) upper bound on the signal's detectability.

\subsection{Bispectra}

\begin{figure}[t]
    \centering
    \includegraphics[width=0.56\linewidth]{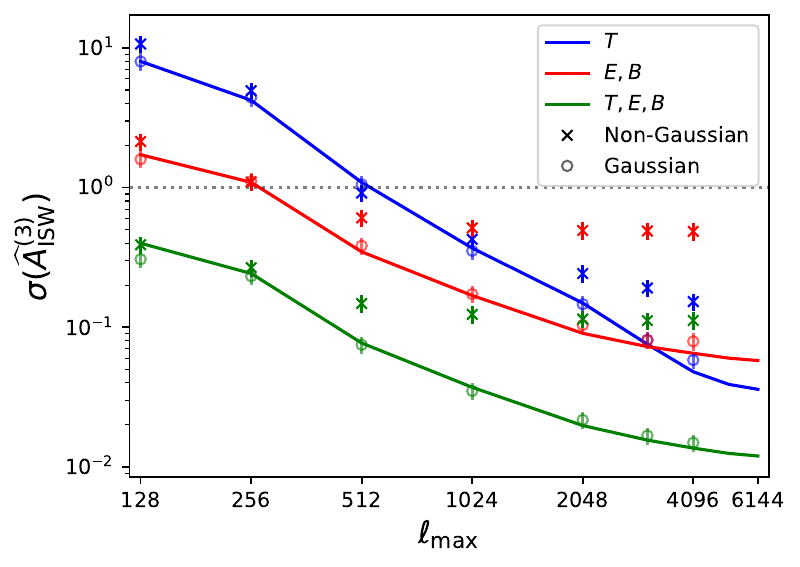}
    \includegraphics[width=0.43\linewidth]{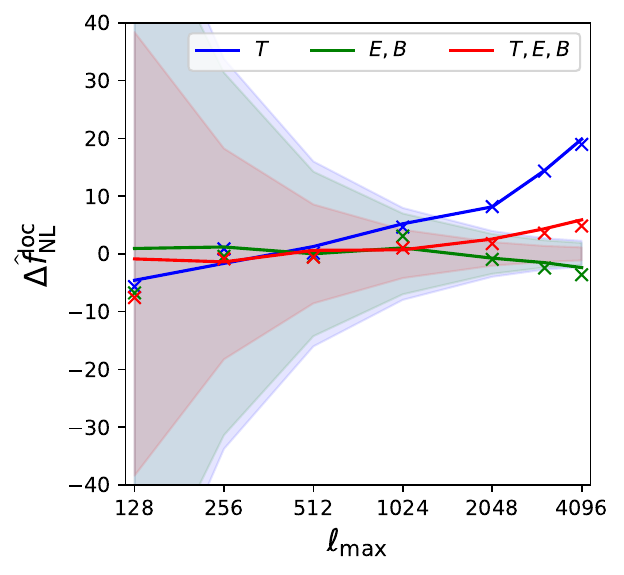}
    \caption{\textbf{Left}: Forecasted constraints on the ISW-lensing bispectrum from an ideal cosmic-variance-limited experiment including all modes up to $\ell_{\rm max}$. The solid curves show the idealized errors obtained from the numerical Fisher matrix, whilst the crosses (circles) indicate the empirical variances obtained from $50$ non-Gaussian (Gaussian) simulations up to $\ell_{\rm max}=4096$. The close agreement between Gaussian simulations and the Fisher forecast imply that our estimators are optimal in the idealized regime; the plateau seen in the non-Gaussian errorbars indicate that lensing-induced non-Gaussianity eventually dominates the covariance (as found in \citep{Lewis:2011fk}). \textbf{Right}: Theoretical (lines) and empirical (crosses) bias on $\fnl$ induced by the ISW-lensing bispectra shown in the left panel. The solid curves indicate the $1\sigma$ Fisher error on $\fnl$ (which are consistent with the empirical lensed variances). We find significant biases in the small-scale temperature-only constraints, reaching $8\sigma$ at $\ell_{\rm max}=2048$.}
    \label{fig: bis-forecast}
\end{figure}

\noindent The left panel of Fig.\,\ref{fig: bis-forecast} shows the forecasted $1\sigma$ errorbar on the ISW-lensing bispectrum amplitude, assuming a cosmic-variance-limited experiment up to $\ell_{\rm max}=6144$, as above. In this idealized limit, we obtain a $3\sigma$ detection of $A_{\rm ISW}^{(3)}$ by $\ell_{\rm max}\approx 1100$ in temperature, $500$ in polarization and $200$ in a joint analysis. Unlike for \textit{Planck} (Fig.\,\ref{fig: bis-results}), the large-scale constraints are dominated by polarization; this occurs since the $B$-modes are limited only by the small gravitational lensing signal, rather than a large noise floor. We find strong dependence of $\sigma(A_{\rm ISW}^{(3)})$ on $\ell_{\rm max}$, particularly for temperature-only analyses, with the precise form related to the shapes of $C_\ell^{T\phi}$ and $C_\ell^{E\phi}$, as well as the lensing reconstruction noise. 

Comparing theoretical and empirical errorbars, we find good agreement on large scales (as expected), but a plateau in the latter for $\ell\gtrsim 1000$ (primarily in polarization).\footnote{At $\ell_{\rm max}=2048$ our results are consistent with the \textit{Planck} constraints, given the reduced sky fraction ($f_{\rm sky}\approx 0.7$) of the latter and noise limitations.} This strongly limits the signal-to-noise: whilst the Fisher forecast indicates that $A_{\rm ISW}^{(3)}$ should be detected at $12\sigma$ ($60\sigma$) at $\ell_{\rm max}=4096$ excluding (including) polarization, we find only a $5\sigma$ ($9\sigma$) detection in the FFP10 simulations. This indicates either that (a) the estimators are not accurate on small scales or (b) non-Gaussian contributions to the covariance matrix are large, \resub{and thus the Cram\'er-Rao bound \eqref{eq: cramer-rao} is not saturated}. To test this, we additionally analyze 50 simulations with the same (lensed) power spectra as the FFP10 suite, but without lensing non-Gaussianity. In this case, we find excellent agreement with the Fisher forecast, indicating that the aforementioned plateau is sourced by lensing non-Gaussianity (matching the conclusions of \citep{Lewis:2011fk}). This could be ameliorated by a more complex estimator, involving, for example, iterative reconstruction \citep[e.g.,][]{Hirata:2003ka,Carron:2017mqf,Belkner:2023duz}.

As noted above, the ISW-lensing bispectrum induces bias on primordial non-Gaussianity parameters, principally $\fnl$. In the right panel of Fig.\,\ref{fig: bis-forecast} we show the induced bias for a cosmic-variance-limited experiment, obtained through a joint analysis of $\fnl$ and $A_{\rm ISW}^{(3)}$ via \eqref{eq: template-biases} (which does not assume the optimality).\footnote{For this purpose, we add the $\fnl$ template to the \polyspec code. The corresponding estimator is derived analogously to $\gnl$ and is equivalent to that of \citep{2011MNRAS.417....2S,Komatsu:2003iq}. A full presentation of the \polyspec primordial bispectrum estimators might occur in future work.} On large-scales, any bias is swamped by the cosmic-variance errorbars; however, we observe a significant bias in temperature-modes from $\ell_{\rm max}\gtrsim 1000$, which is consistent between the non-Gaussian simulations and theory. This occurs due to the tightening of $\fnl$ constraints with scale (with $\sigma(\fnl)\sim \ell_{\rm max}^{-1}$ \citep[cf.][]{Kalaja:2020mkq}), and the large overlap of ISW and local templates, which reaches $40\%$ by $\ell_{\rm max}=4096$. For polarization, we find no significant bias due to the different physical signatures.  These conclusions are consistent with previous results \citep[e.g.,][]{2011MNRAS.417....2S,Planck:2019kim,Planck:2015zfm,Hill:2018ypf,Kim:2013nea} and highlight the importance of accounting for such biases in future surveys. When performing a joint analysis of $\fnl$ and $A_{\rm ISW}^{(3)}$, the errorbars on $\fnl$ are relatively insensitive to lensing non-Gaussianity: at $\ell_{\rm max}=4096$ the theoretical and empirical temperature-plus-polarization errors are consistent within $30\%$, matching \citep{Coulton:2019odk}.

\subsection{Trispectra}
\begin{figure}
    \centering
    \includegraphics[width=0.49\linewidth]{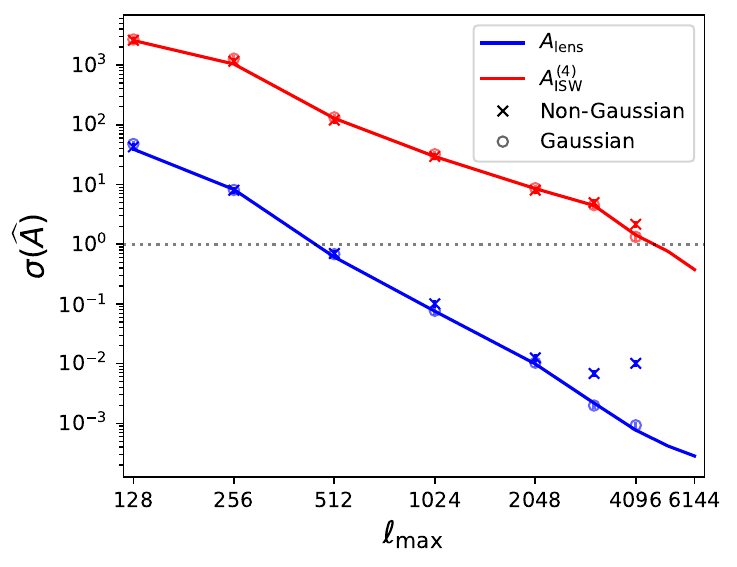}
    \includegraphics[width=0.49\linewidth]{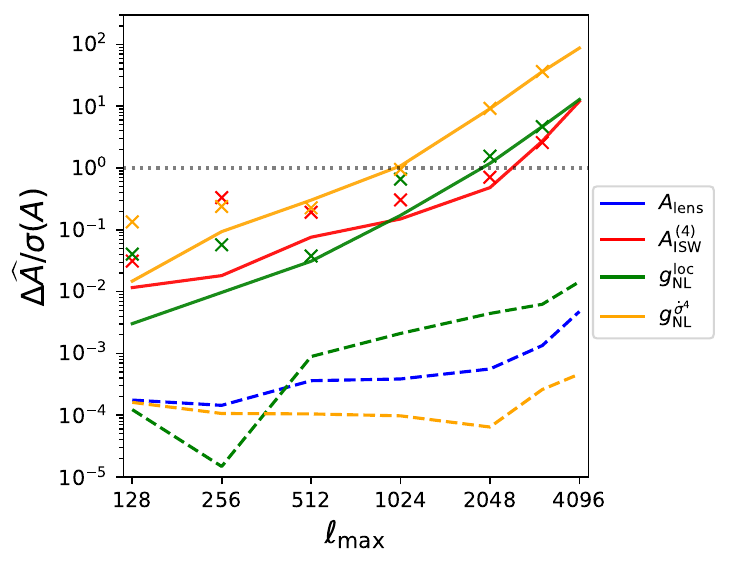}
    \caption{\textbf{Left}: Forecasted constraints on the lensing (blue) and ISW-lensing (red) trispectrum from an ideal cosmic-variance-limited experiment, including all temperature modes up to $\ell_{\rm max}$, restricting to temperature. As in Fig.\,\ref{fig: bis-forecast}, the lines indicate theoretical forecasts whilst the crosses (circles) show results from non-Gaussian (Gaussian) simulations. Whilst the lensing trispectrum will be detectable at high signal-to-noise, measuring the fiducial ISW-lensing correlations in future data will be challenging. \textbf{Right}: Fractional biases on four trispectrum amplitudes induced by lensing (solid lines) and ISW-lensing (dashed lines), as in Fig.\,\ref{fig: tris-results}. The crosses indicate the biases measured from cosmic-variance-limited simulations, which up to noise, are consistent with theory. Though lensing induces large biases, ISW-lensing effects remain a small fraction of a sigma.}
    \label{fig: tris-forecast}
\end{figure}

\noindent In Fig.\,\ref{fig: tris-forecast} we perform a similar analysis for the late-time trispectra, focusing on the temperature sector, as in \S\ref{sec: current-data}. For lensing, we forecast strong detections for all $\ell_{\rm max}\gtrsim 500$ (matching many forecasts \citep[e.g.,][]{Belkner:2023duz}), around $100\sigma$ at $\ell_{\rm max}=2048$ and almost $1000\sigma$ at $\ell_{\rm max}=6144$. As for $A_{\rm ISW}^{(3)}$ there is some evidence for a plateau in the non-Gaussian errorbars at large $\ell_{\rm max}$, motivating alternative estimators such as iterative schemes, though this is less significant than before, particularly since we do not include polarization. For the ISW-lensing trispectrum, the situation remains bleak: even in an ideal cosmic-variance-limited experiment at $\ell_{\rm max}=6144$, we detect $A_{\rm ISW}^{(4)}$ only at $2.7\sigma$. That said, these constraints are not strongly limited by likelihood non-Gaussianity (as evidenced by the excellent agreement between theoretical and empirical errors, unlike for $A_{\rm ISW}^{(3)}$), and could be significantly enhanced with the addition of polarization. An optimist's summary, therefore, is that a future low-noise experiment could detect the ISW-lensing trispectrum at modest significance.

In the right panel of Fig.\,\ref{fig: tris-forecast}, we forecast the induced biases on primordial non-Gaussianity parameters (considering $\gnl$ and $\gnldotdot$, which showed the strongest responses to ISW-lensing and lensing respectively in Fig.\,\ref{fig: tris-forecast}). A non-zero value of $A_{\rm lens}$ leads to significant bias in both $\gnl$ and $\gnldotdot$ as well as $A_{\rm ISW}^{(4)}$ (all of which are consistent with the values obtained from simulations) -- this is unsurprising given its large signal-to-noise. In contrast, we forecast negligible ($<0.15\sigma$) biases on both the primordial templates and $A_{\rm lens}$ from ISW-lensing correlations. This is an important conclusion: the ISW-lensing contractions discussed in this work do not need to be accounted for in any previous or (near-)future lensing or primordial four-point studies \citep[e.g.,][]{2015arXiv150200635S,Philcox4pt3,Carron:2022eyg}.

\section{Discussion}\label{sec: summary}
\noindent Weak gravitational lensing and the integrated Sachs-Wolfe effect are sourced by the same potentials; as such, their effects on the CMB are correlated. This sources non-Gaussianity in the CMB, creating an ISW-induced bispectrum, trispectrum, and beyond. Whilst the three-point function has been shown to be a valuable probe of late-time physics such as dark energy and modified gravity \citep[e.g.,][]{Hu:2012td,Yamauchi:2013pna,Munshi:2014tua,Kable:2021yws,Jain:2007yk,Giannantonio:2013kqa,Calabrese:2009tt,Chudaykin:2025gdn}, it is \textit{a priori} unclear whether the same holds for the four-point function. In this work, we have endeavored to elucidate this question.

By constructing a suite of minimum-variance estimators and applying them to current and simulated data, we have obtained bounds on both the ISW-lensing bispectrum and trispectrum, and forecasted the detectability in future cosmic-variance-limited experiments. Whilst we find strong detections of the three-point amplitude, consistent with both previous works and the fiducial model \citep[e.g.,][]{Planck:2015mym,Carron:2022eyg}, we do not detect the four-point function, finding an errorbar $15\times$ larger than the fiducial signal. Future high-resolution experiments could yield sharp constraints on the three-point function and may (just) detect the four-point function -- to maximize the information content on the former, however, we will likely require iterative lensing estimators \citep[e.g.,][]{Belkner:2023duz}. 

As discussed in \S\ref{sec: flat-sky}, the small signal-to-noise of the ISW-lensing trispectrum is due to an almost perfect cancellation between an exchange- and a contact-type term, which occur at first- and second-order in the lensing expansion respectively. Even in a futuristic cosmic-variance-limited experiment at $\ell_{\rm max}\sim 6000$ the signal will be difficult to detect, though the prospects are somewhat enhanced when polarization is included. This implies that constraints on modified gravity from the ISW signal present in CMB auto-correlations are essentially saturated by the three-point function. Furthermore, it seems unlikely that one can use the quadratic estimator of \eqref{eq: quad-ISW} to meaningfully constrain the unlensed temperature field from the lensed data. In both the above cases, however, more information can be extracted by considering cross-spectra with late-time tracers such as galaxy surveys. The small amplitude of the ISW-lensing trispectrum implies negligible biases on both current and future measurements of lensing and primordial non-Gaussianity. This conclusion is not specific to ISW: rather, it would hold for any effect that induces correlations between the lensing potential and the CMB fluctuations, provided that it dominates on large scales.

The ISW-lensing trispectrum discussed in this work arises at second-order in the lensing potential, $\phi$. At the same order, additional effects arise, such as post-Born effects, lensing rotation, and late-time non-Gaussianity \citep[e.g.,][]{Marozzi:2016qxl,Marozzi:2016uob,Pratten:2016dsm}, all of which produce analogous trispectra. Notably, these are not subject to the same cancellations as the fiducial signal (since they are not present at first order); in principle, they could thus be measurable in the four-point function. Whilst a detailed analysis of such effects is beyond the scope of this work, we argue in Appendix \ref{app: post-born} that these give subdominant contributions to the ISW-lensing trispectrum, due to the large-scale restrictions inherent in ISW-lensing cross-correlations.

A number of other late-time effects can also source CMB trispectra. In particular, distortions induced by the cosmic infrared background and the thermal Sunyaev-Zel'dovich source temperature-lensing correlations analogous to those discussed in this work. Whilst there are detailed analytic and numerical forecasts for the induced bispectra \citep{Hill:2018ypf,Coulton:2022wln}, the impact of these effects on the trispectrum has yet to be considered. Notably, the relevant cross-spectra are not restricted to large-scales, thus we do not expect the strong cancellations between first- and second-order terms seen in this work. However, many of these effects distort the frequency spectrum of the CMB and can thus be removed by multi-frequency cleaning. We leave a more thorough treatment of such signals to future work. 

To finish, let us return to the questions posed in the introduction. Is the ISW-lensing trispectrum detectable? Can it bias measurements of CMB lensing? Can it bias measurements of primordial non-Gaussianity? No, no, and no.

\vskip 8pt
\textit{Note Added}: Whilst finalizing this work, \citep{Jung:2025nss} appeared on arXiv, which includes an analysis of the ISW-lensing bispectrum using \textit{Planck} PR4 temperature- and polarization-data. Whilst the methodology differs (with the former work using binned estimators), our results are comparable. 

\acknowledgments
{\small
\begingroup
\hypersetup{hidelinks}
\noindent 
We are highly grateful to Anthony Challinor for insightful comments on the manuscript, \resub{and thank the anonymous referee for comments}. OHEP is a Junior Fellow of the Simons Society of Fellows, and thanks \href{https://www.flickr.com/photos/198816819@N07/54386115227/}{Llouis Armstrong} for jazzy grooves. JCH acknowledges support from the Sloan Foundation and the Simons Foundation.
\endgroup
The computations in this work were run at facilities supported by the Scientific Computing Core at the Flatiron Institute, a division of the Simons Foundation.
}

\appendix

\section{Higher-Order Effects}\label{app: post-born}

\noindent Throughout this work, we have computed results assuming the \textit{Born approximation}, \textit{i.e.}\ that photons travel along along unperturbed geodesics. Violations of this assumption occur at second-order in the lensing potential and above, providing an additional source of ISW-lensing cross-correlations. As discussed in \citep{Pratten:2016dsm,Marozzi:2016qxl,Marozzi:2016uob}, this modifies the remapping equation to
\beq\label{eq: lensing-flat-sky-higher-order}
    \tilde{T}(\btheta,\chi) \equiv T(\btheta+\delta\btheta(\btheta,\chi),\chi) = T(\btheta,\chi)+\nabla_iT(\btheta,\chi)\delta\btheta^i(\btheta,\chi)+\frac{1}{2}\nabla_i\nabla_j T(\btheta,\chi)\delta\btheta^i(\btheta,\chi)\delta\btheta^j(\btheta,\chi)+\cdots,
\eeq
for a source at distance $\chi$ in the flat-sky limit at leading-order, where $\delta\btheta^i$ is the non-linear change in the photon angle. Up to second-order in the Weyl potential, $\Phi$, 
\beq
    \delta\btheta^i(\btheta,\chi) = -2\int_0^{\chi}d\chi'\frac{\chi-\chi'}{\chi\chi'}\nabla^i\Phi(\btheta,\chi') + 4\int_0^{\chi}d\chi'\frac{\chi-\chi'}{\chi\chi'}\int_0^{\chi'}d\chi''\frac{\chi'-\chi''}{\chi'\chi''}\nabla_{ij}\Phi(\btheta,\chi')\nabla^j\Phi(\btheta,\chi'')+\cdots,
\eeq
where the first term is equal to $\nabla_i\phi(\btheta,\chi)$. 
This sources a new contribution to the temperature field (which includes curl lensing contributions):
\beq
    \tilde{T}(\vl,\chi) &\supset & 4\int_0^{\chi}d\chi'\frac{\chi-\chi'}{\chi\chi'}\int_0^{\chi'}d\chi''\frac{\chi'-\chi''}{\chi'\chi''}\int_{\vL\vL'}[\vL'\cdot(\vL-\vL')][\vL'\cdot(\vl-\vL)]\Phi(\vL',\chi')\Phi(\vL-\vL',\chi'')T(\vl-\vL,\chi)
\eeq
\citep{Marozzi:2016qxl}, and thus the flat-sky trispectrum:
\beq\label{eq: flat-sky-born}
    \av{\tilde{T}(\vl_1)\tilde{T}(\vl_2)\tilde{T}(\vl_3)\tilde{T}(\vl_4)}'_c &\supset& -2[\vl_3\cdot\vl_4][\vl_3\cdot\vl_2]C_{l_2}^{TT}\int_0^{\chi_\star}d\chi\frac{\chi_\star-\chi}{\chi_\star\chi}\left[-2\int_0^{\chi}d\chi'\frac{\chi-\chi'}{\chi\chi'}C_{l_4}^{T\Phi}(\chi')\right]C_{l_3}^{T\Phi}(\chi)\nonumber\\
    &&\,+\,\text{23 perms.},
\eeq
where we note that $\av{T(\vl,\chi)T^*(\vl)}$ peaks at $\chi\approx\chi_\star$, as before. Here, the term in square brackets is equal to the ISW-lensing power spectrum $C_{l_4}^{T\phi}$ evaluated at conformal distance $\chi$. This is similar to the ISW-IV shape, but features an additional $\chi$ integral, which couples two legs of the trispectrum (due to the multi-plane lensing). As in the Born approximation, the trispectrum specified by \eqref{eq: flat-sky-born} is dominated by the doubly-squeezed regime with $l_3,l_4\ll l_1,l_2$ (and permutations thereof). In this limit, the leading-order term vanishes after symmetrization, since $[\vl_3\cdot\vl_2]\approx -[\vl_3\cdot\vl_1]$. Combining this cancellation with the inherent suppression of post-Born effects, we expect that the above trispectrum will be significantly smaller than that of ISW-I or ISW-IV. 

\begin{figure}
    \centering
    \includegraphics[width=0.9\linewidth]{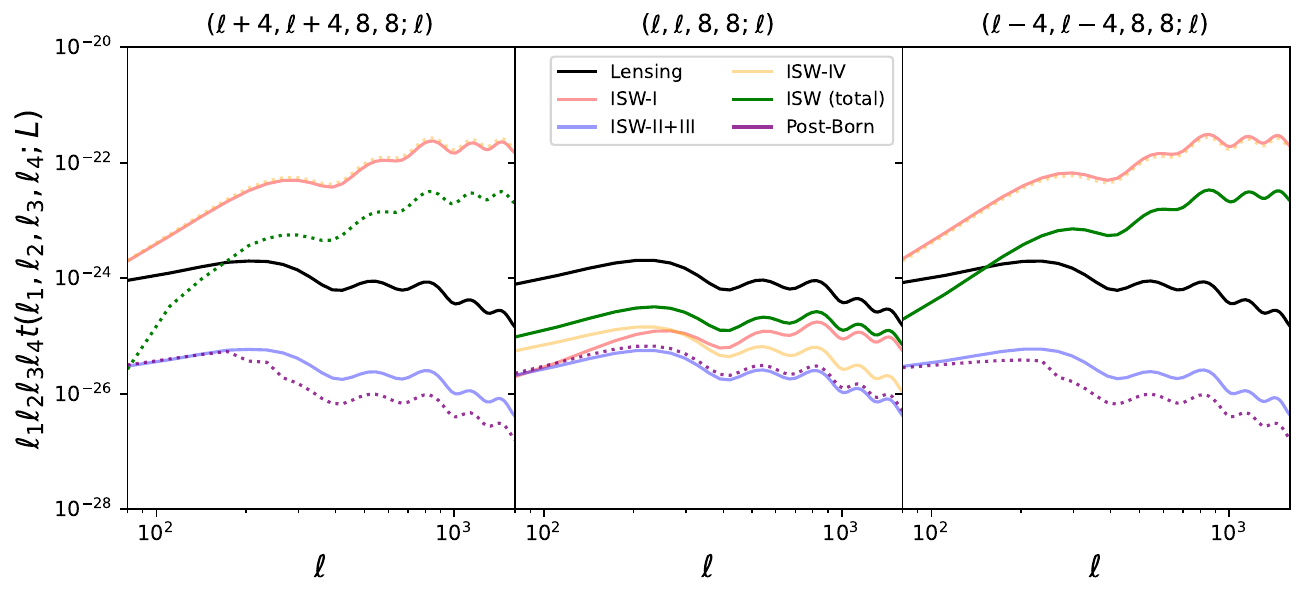}
    \caption{Contributions to the flat-sky trispectrum, $t(l_1,l_2,l_3,l_4;L) = \av{T(\vl_1)T(\vl_2)T(\vl_3)T(\vl_4)}_c$ induced by lensing (black), fiducial ISW-lensing effects (green) and post-Born ISW-lensing effects (purple). We additionally show the individual ISW contributions in red, blue and yellow, and indicate negative contributions by dotted lines. The three panels show different choices of $(l_1,l_2,l_3,l_4,L)$ (with $L\equiv|\vl_1+\vl_3|$), restricting to the doubly-squeezed regime in all cases. Despite the strong cancellation between the fiducial ISW contributions (particularly ISW-I and ISW-IV), the post-Born corrections are highly subdominant to the total signal.}
    \label{fig: post-born}
\end{figure}

To assess whether the post-Born trispectrum is comparable to the \textit{combined} fiducial signal (given the cancellations discussed in \S\ref{sec: flat-sky}), we plot the two signals in Fig.\,\ref{fig: post-born}, computing the former using a finely sampled numerical integral, with transfer functions obtained from \textsc{camb}. In the doubly-squeezed limit, the shape of the Born trispectrum depends strongly on the angle between the long and short modes, with the combined spectra changing sign around $l_1=l_2=L$ (with $\vL\cdot\vl\to 0$). In all cases, the post-Born corrections represent a very small fraction of the signal -- this also holds for non-squeezed configurations and the collapsed limit. As such, we conclude that the Born approximation is appropriate for the purposes of this study.

Another contribution to the ISW-lensing four-point function is sourced by gravitational non-Gaussianity in the lensing potential, $\phi$ \citep[e.g.,][]{Marozzi:2016uob}. At leading-order, this sources the following trispectrum
\beq\label{eq: grav-isw-trispectra}
    \av{\tilde{T}(\vl_1)\tilde{T}(\vl_2)\tilde{T}(\vl_3)\tilde{T}(\vl_4)}'_c &\supset& \int_{\vL'}(2\pi)^2\delta_{\rm D}(\vl_1+\vl_2-\vL')(\vl_2\cdot\vL)C_{l_2}^{TT}\av{\phi(\vL')T(\vl_3)T(\vl_4)} + \text{11 perms.}
\eeq
involving a potential-temperature bispectrum, $\av{\phi TT}$. From the Poisson equation, $\phi$ traces the inverse Laplacian of the matter density $\delta$, which contains a term quadratic in the linear density field, $\delta_L$, at second order in perturbation theory. Due to the ISW effect, the linear density field correlates with the unlensed temperature field, leading to a trispectrum contribution with the schematic form $\av{TT}\nabla^{-2}[\av{\delta_L T}\av{\delta_L T}]$. Whilst a detailed computation of this effect is beyond the scope of this work (though possible via the methods of \citep{Marozzi:2016qxl,Marozzi:2016uob}), we can gain some understanding of its shape and amplitude by inspecting \eqref{eq: grav-isw-trispectra}. Due to the ISW contributions, this must peak at low $l_3,l_4$, restricting us to the doubly-squeezed limit as before. Similarly to the post-Born trispectrum, the leading-order term cancels after symmetrizing over $\vl_1\leftrightarrow\vl_2$, considerably suppressing the signal. Moreover, non-linear corrections to $\phi$ are important only on small scales, \textit{i.e.}\ large $L'$; however, momentum conservation forces $L'$ to be small thus $\phi$ to be in the quasi-linear regime. As a result, we expect non-linear corrections to the ISW-lensing trispectrum to be highly subdominant and ignore them in the main analyses of this work.

\section{Extension to Polarization}\label{app: pol}
\noindent In this appendix, we generalize the correlators and estimators described in the main text to include polarization. First, we compute the perturbations to the polarization tensor, $\P_{ij}(\hn)$, induced by the lensing remapping at second-order, before computing the correlators, and presenting the generalized bispectrum and trispectrum estimators.

\subsection{Lensing Non-Gaussianity}
\noindent As discussed in \citep{Okamoto:2003zw}, it is convenient to work with the spin-$\pm2$ field ${}_{\pm 2}A(\hn)$, defined via $\P_{ij} = \bar{\vec{m}}_i\bar{\vec{m}}_j{}_{+2}A(\hn)+\vec{m}_i\vec{m}_j{}_{-2}A(\hn)$, where $\vec{m},\bar{\vec{m}}$ are null vectors on the sphere, satisfying $\vec{m}\cdot\vec{m} = \bar{\vec{m}}\cdot\bar{\vec{m}} = 0,\,\vec{m}\cdot\bar{\vec{m}}=1$. In the presence of lensing, a spin-$\pm2$ field sourced at distance $\chi$ transforms as
\beq
    {}_{\pm2}\tilde{A}(\hn,\chi) \equiv {}_{\pm2}A(\hn+\nabla\phi(\hn,\chi),\chi) &=& {}_{\pm2}A(\hn,\chi) + D^k[{}_{\pm2}A(\hn,\chi)]D_k\phi(\hn,\chi)\\\nonumber
    &&\,+\,\frac{1}{2}D^kD^l[{}_{\pm2}A(\hn,\chi)]D_k\phi(\hn,\chi)D_l\phi(\hn,\chi)+\cdots,
\eeq
where $D^k$ are gradients in the spin-weighted representation \citep{Okamoto:2003zw}. Expanding in spherical harmonics, we find
\beq
    {}_{\pm2}A_{\ell m}(\chi) &\to& {}_{\pm2}A_{\ell m}(\chi) + \sum_{LM\ell'm'}{}_{\pm2}A_{\ell'm'}(\chi)\phi_{LM}(\chi)\int d\hn\,{}_{\pm2}Y^*_{\ell m}(\hn)D^k[{}_{\pm2}Y_{\ell'm'}(\hn)]D_k[Y_{LM}(\hn)]\\\nonumber
    &&\,+\,\frac{1}{2}\sum_{LML'M'\ell'm'}{}_{\pm2}A_{\ell'm'}(\chi)\phi_{LM}(\chi)\phi_{L'M'}(\chi)\\\nonumber
    &&\qquad\qquad\,\times\,\int d\hn\,{}_{\pm2}Y^*_{\ell m}(\hn)D^kD^l[{}_{\pm2}Y_{\ell'm'}(\hn)]D_k[Y_{LM}(\hn)]D_l[Y_{L'M'}(\hn)]+\cdots
\eeq
with ${}_{\pm 2}A_{\ell m} = a^E_{\ell m}\pm ia^B_{\ell m}$. As in \eqref{eq: lensing-full-sky-harmonic}, we can define generalized coupling kernels such that
\beq\label{eq: lensing-full-sky-pol}
    \tilde{a}_{\ell m}^X(\chi)&=& a_{\ell m}^X(\chi) + (-1)^m\sum_{\ell'm'LM}\mathcal{I}^{\ell mX}_{\ell'm'X'LM}a_{\ell'm'}^{X'}(\chi)\phi_{LM}(\chi)\\\nonumber
    &&\,+\,\frac{1}{2}(-1)^{m}\sum_{\ell'm'X'LML'M'}\mathcal{J}^{\ell mX}_{\ell'm'X'LML'M'}a^{X'}_{\ell'm'}(\chi)\phi_{LM}(\chi)\phi_{L'M'}(\chi)+\cdots,
\eeq
for $X\in\{T,E,B\}$ with spin $s_X\in\{0,2,2\}$. Noting that
\beq
    D_k[{}_sY_{\ell m}(\hn)] &=& -\frac{1}{\sqrt{2}}\left[\sqrt{(\ell-s)(\ell+s+1)}{}_{s+1}Y_{\ell m}(\hn)\bar{\vec m}_k-\sqrt{(\ell+s)(\ell-s+1)}{}_{s-1}Y_{\ell m}(\hn)\vec{m}_k\right]
\eeq
\citep{Okamoto:2003zw}, the quadratic kernel is given by
\beq\label{eq: quadratic-kernel-pol}
    \mathcal{I}^{\ell m X}_{\ell'm'X'LM}&\equiv &-\frac{1}{2}\left[\epsilon_{\ell\ell'L}\delta_{\rm K}^{X'X}-\beta_{\ell\ell'L}\delta_{\rm K}^{X'\bar{X}}\right]\sqrt{L(L+1)}\sum_{\lambda=\pm1}\sqrt{(\ell'+\lambda s_X)(\ell'-\lambda s_X+1)}\\\nonumber
    &&\,\times\,\int d\hn\,{}_{-s_X}Y^*_{\ell m}(\hn){}_{s_X-\lambda}Y^*_{\ell'-m'}(\hn) {}_{\lambda}Y^*_{L-M}(\hn)
\eeq
where 
\beq
    2\epsilon_{\ell_1\ell_2\ell_3} = 1+(-1)^{\ell_1+\ell_2+\ell_3}, \qquad 2i\beta_{\ell_1\ell_2\ell_3} = 1-(-1)^{\ell_1+\ell_2+\ell_3}
\eeq
with $a_{\ell m}^{\bar{E}} = -a_{\ell m}^B, a_{\ell m}^{\bar{B}} = a_{\ell m}^E$ and $a_{\ell m}^{\bar{T}}=0$. This is derived by differencing the expression for ${}_{|s|}A$ and ${}_{-|s|}A$ and noting that ${}_sA_{\ell m}(-\hn) = (-1)^\ell{}_{-s}A_{\ell m}(\hn)$. This is equivalent to the form derived in \citep{Okamoto:2003zw} and satisfies $\left(\mathcal{I}^{\ell(-m)X}_{\ell'(-m')X'\ell''(-m'')}\right)^* =\mathcal{I}^{\ell mX}_{\ell'm'X'\ell''m''}$. Similarly, the cubic kernel is defined as
\beq\label{eq: cubic-kernel-pol}
    \mathcal{J}^{\ell m X}_{\ell'm'X'LML'M'} &=& \frac{1}{4}\left[\epsilon_{\ell\ell'LL'}\delta_{\rm K}^{XX'}-\beta_{\ell\ell'LL'}\delta_{\rm K}^{\bar{X}X'}\right]\sqrt{L(L+1)L'(L'+1)}\int d\hn\,{}_{-s_X}Y^*_{\ell m}(\hn)\sum_{\lambda=\pm1}{}_{\lambda}Y^*_{L-M}(\hn)\\\nonumber
    &&\,\times\,\bigg\{\sqrt{(\ell'+\lambda s_X)(\ell'-\lambda s_X+1)(\ell'+\lambda s_X-1)(\ell'-\lambda s_X+2)}{}_{\lambda}Y^*_{L'-M'}(\hn){}_{s_X-2\lambda}Y^*_{\ell'-m'}(\hn)\\\nonumber
    &&\,\qquad\,+(\ell'+\lambda s_X)(\ell'-\lambda s_X+1){}_{-\lambda}Y^*_{L'-M'}(\hn){}_{s_X}Y^*_{\ell'-m'}(\hn)\bigg\}.
\eeq
where $\epsilon,\beta$ are analogous to before. These forms reduce to the temperature-only kernels of \eqref{eq: quadratic-kernel}\,\&\,\eqref{eq: cubic-kernel} for $X=X'=T$.

\subsection{Correlators}
\noindent As in \S\ref{subsec: correlators}, the leading-order bispectrum is sourced by one second-order field and two leading-order fields in \eqref{eq: lensing-full-sky-pol}. This gives the three-point function
\beq\label{eq: bis-full-pol}
    \av{a_{\ell_1m_1}^{X_1}a_{\ell_2m_2}^{X_2}a_{\ell_3m_3}^{X_3}}_c &\supset& \sum_Y\mathcal{I}^{\ell_1m_1X_1}_{\ell_2(-m_2)Y\ell_3(-m_3)}C_{\ell_2}^{X_2Y}C_{\ell_3}^{X_3\phi} + \text{5 perms.}\\\nonumber
    &=&\sqrt{\frac{(2\ell_1+1)(2\ell_2+1)(2\ell_3+1)}{4\pi}}\tj{\ell_1}{\ell_2}{\ell_3}{m_1}{m_2}{m_3}\tj{\ell_1}{\ell_2}{\ell_3}{s_{X_1}}{-s_{X_1}}{0}\\\nonumber
    &&\,\times\,\frac{1}{2}\left(\ell_2(\ell_2+1)+\ell_3(\ell_3+1)-\ell_1(\ell_1+1)\right)\left[\epsilon_{\ell_1\ell_2\ell_3}C_{\ell_2}^{X_1X_2}-\beta_{\ell_1\ell_2\ell_3}C_{\ell_2}^{\bar{X}_1X_2}\right]C_{\ell_3}^{X_3\phi} + \text{5 perms},
\eeq
where the square bracket is equal to $C_{\ell_2}^{X_1X_2}$ if $\ell_1+\ell_2+\ell_3$ is even and $iC_{\ell_2}^{\bar{X_1}X_2}$ else (explicitly breaking parity). Here, we have noted that the $\chi$ integrals are dominated by $\chi\approx\chi_\star$, as in \S\ref{sec: full-sky}. Assuming that $\phi$ correlates only with $T$- and $E$-modes, this involves all combinations of temperature and polarization except $BBB$ \citep[e.g.,][]{Hu:2001kj,Okamoto:2003zw}.

With the addition of polarization, the first-order lensing correction is no longer symmetric under interchange of the lensing potential and the unlensed field, which results in a more complex trispectrum than for the temperature-only case. This can be seen from the quadratic estimators:
\beq\label{eq: lens-isw-quad-correlators-pol}
    \av{a^{X_1}_{\ell_1m_1}a^{X_3}_{\ell_3m_3}}_{\rm unlensed} &=& \sum_{LMY}(-1)^{M}\mathcal{I}^{\ell_1m_1X_1}_{\ell_3(-m_3)YLM}C_{\ell_3}^{X_3Y}\phi_{LM} + (1\leftrightarrow3)\\\nonumber
    \av{a^{X_1}_{\ell_1m_1}a^{X_3}_{\ell_3m_3}}_{\rm ISW}&=& \sum_{LMY}(-1)^{M}\mathcal{I}^{\ell_1m_1X_1}_{LMY\ell_3(-m_3)}\int_0^{\chi_\star}d\chi\,C_{\ell_3}^{\phi X_3}(\chi)a_{LM}^{Y}(\chi) + (1\leftrightarrow3),
\eeq
where the second term describes the exchange of a general field $a_{\ell m}^{Y}(\chi)$ where $Y\in\{T,E,B\}$, and we note the indexing asymmetries in $\mathcal{I}$. The full exchange trispectrum is given by:
\beq\label{eq: tris-ex-full-sky-pol}
    \av{a_{\ell_1m_1}^{X_1}a_{\ell_2m_2}^{X_2}a_{\ell_3m_3}^{X_3}a_{\ell_4m_4}^{X_4}}_c&\supset&\quad\sum_{LMYY'}(-1)^M\mathcal{I}^{\ell_1m_1X_1}_{\ell_3(-m_3)YLM}\mathcal{I}^{\ell_2m_2X_2}_{\ell_4(-m_4)Y'L(-M)}C_{\ell_3}^{X_3Y}C_{\ell_4}^{X_4Y'}C_L^{\phi\phi}\\\nonumber
    &&\,+\,\sum_{LMYY'}(-1)^{M}\mathcal{I}^{\ell_1m_1X_1}_{LMY\ell_3(-m_3)}\mathcal{I}^{\ell_2m_2X_2}_{L(-M)Y'\ell_4(-m_4)}C_{\ell_3}^{X_3\phi}C_{\ell_4}^{X_4\phi}C_L^{YY'}\\\nonumber
    &&\,+\,\sum_{LMYY'}(-1)^M\mathcal{I}^{\ell_1m_1X_1}_{LMY\ell_3(-m_3)}\mathcal{I}^{\ell_2m_2X_2}_{\ell_4(-m_4)Y'L(-M)}C_{\ell_4}^{X_4Y'}\int_0^{\chi_\star}d\chi\,C_{\ell_3}^{X_3\phi}(\chi)C_L^{\phi Y}(\chi)\\\nonumber
    &&\,+\,\sum_{LMYY'}(-1)^{M}\mathcal{I}^{\ell_1m_1X_1}_{\ell_3(-m_3)YLM}\mathcal{I}^{\ell_2m_2X_2}_{L(-M)Y'\ell_4(-m_4)}C_{\ell_3}^{X_3Y}\int_0^{\chi_\star}d\chi\,C_{\ell_4}^{X_4\phi}(\chi)C_L^{\phi Y'}(\chi)+\text{11 perms.},
\eeq
encoding lensing, ISW-I, ISW-II and ISW-III as before, and simplifying the $\chi$ integrals where possible, noting that the ISW-II and ISW-III contributions can be sourced by lensing of both the ISW signal and reionization anisotropies. The contact trispectrum (ISW-IV) is analogous to the bispectrum with
\beq\label{eq: tris-con-full-sky-pol}
    \av{a_{\ell_1m_1}^{X_1}a_{\ell_2m_2}^{X_2}a_{\ell_3m_3}^{X_3}a_{\ell_4m_4}^{X_4}}_c &\supset& \frac{1}{2}\sum_{Y}\mathcal{J}^{\ell_1m_1X_1}_{\ell_2(-m_2)Y\ell_3(-m_3)\ell_4(-m_4)}C_{\ell_2}^{X_2Y}C_{\ell_3}^{X_3\phi}C_{\ell_4}^{X_4\phi}+\text{23 perms.}
\eeq
Both trispectra include parity-odd contributions with odd $\ell_1+\ell_2+\ell_3+\ell_4$. Note that the ISW-lensing trispectra can contain at most two $B$-modes (except for the cross-terms, which can contain three).

\subsection{Bispectrum Estimator}
\noindent The polarized bispectrum estimator can be derived similarly to \S\ref{subsec: bspec-estimator}. Starting from \eqref{eq: gen-estimator} and separating out the linear term as in \eqref{eq: b-numerator}, we obtain the cubic estimator
\beq\label{eq: estimator-numerator-bis-pol}
    \widehat{\mathcal{N}}_{A^{(3)}_{\rm ISW}}[\alpha,\beta,\gamma] &=& \frac{1}{6}\sum_{Y}\mathcal{I}^{\ell_1m_1X_1}_{\ell_2(-m_2)Y\ell_3(-m_3)}C_{\ell_3}^{X_3\phi}C_{\ell_2}^{X_2Y}[\Si\alpha]^{X_1*}_{\ell_1m_1}[\Si\beta]^{X_2*}_{\ell_2m_2}[\Si\gamma]^{X_3*}_{\ell_3m_3}+\text{5 perms.}\\\nonumber
    &=&\frac{1}{24}\left[\sum_{\lambda=\pm 1}\int d\hn\,\sum_X\left({}_{s_X}U^{X}[\Si\alpha](\hn){}_{s_X}V_{\lambda}^{X,\rm lens*}[\Si\beta](\hn)\right)V_{\lambda}^{{\rm ISW}}[\Si\gamma](\hn)+\text{c.c.}\right]+\text{5 perms.},
\eeq
after extensive simplification, where we have defined the spin $(s_X,-\lambda,s_X-\lambda)$ maps (including a $\chi$ argument in $V^{\rm ISW}_\lambda$ for later use)
\beq\label{eq: UV-maps-pol}
    &&{}_{s_X}U^X[x](\hn) \equiv \sum_{\ell m}{}_{s_X}Y_{\ell m}(\hn)x^X_{\ell m}, \qquad V^{\rm ISW}_\lambda[x](\hn,\chi) \equiv \sum_{\ell m X}{}_{-\lambda}Y_{\ell m}(\hn)\sqrt{\ell(\ell+1)}C_{\ell}^{X\phi}(\chi)x^{X}_{\ell m}\\\nonumber
    &&{}_{s_X}V^{{\rm lens},X}_\lambda[x](\hn) \equiv \sum_{\ell m Z}{}_{s_X-\lambda}Y_{\ell m}(\hn)\sqrt{(\ell+\lambda s_X)(\ell-\lambda s_X+1)}\left(C_{\ell}^{XZ}-iC_{\ell}^{\bar{X}Z}\right)x^{Z}_{\ell m}
\eeq
which reduce to \eqref{eq: UV-maps-T} for $X=T,s_X=0$. Note that \eqref{eq: estimator-numerator-bis-pol} is explicitly real and can be computed via spherical harmonic transforms (up to spin-three), and a pixel-space sum. Furthermore, it can be rewritten in terms of the quadratic lensing estimator defined below in \eqref{eq: lens-isw-quad-estimators-pol}:
\beq
    \widehat{\mathcal{N}}_{A^{(3)}_{\rm ISW}}[\alpha,\beta,\gamma] &=& \frac{1}{6}\sum_{LMX}\sqrt{L(L+1)}C_L^{X\phi}[\Si\gamma]_{LM}^{X*}\Phi^{\rm lens}_{LM}[\Si\alpha,\Si\beta]+\text{5 perms.},
\eeq
which is simply the cross-spectrum of the reconstructed potential and the CMB $T$- and $E$-modes, analogous to \citep{Carron:2022eum} (which ignored $E\phi$ correlations).

Following a similar method to \S\ref{subsec: bspec-estimator}, we can compute the normalization term. This involves the $Q_{\ell m}^X$ derivatives (adding a polarization index to \eqref{eq: fish3}), which take the lengthy form:
\beq
    Q^X_{\ell m,A^{(3)}_{\rm ISW}}[x,y] &\equiv& \sum_{\ell_2\ell_3m_2m_3X_2X_3}\frac{\partial \av{a^{X}_{\ell m}a^{X_2}_{\ell_2m_2}a^{X_3}_{\ell_3m_3}}_c}{\partial A^{(3)}_{\rm ISW}}x^{X_2*}_{\ell_2m_2}y^{X_3*}_{\ell_3m_3}\\\nonumber
    &=&-\frac{1}{4}\int d\hn\,\left[{}_{s_X}Y^*_{\ell m}(\hn){}_{s_X}F^{(1),X}[x,y](\hn)+{}_{-s_X}Y^*_{\ell m}(\hn){}_{s_X}F^{(1),X*}[x,y](\hn)\right]\\\nonumber
    &&\,+\frac{1}{4}\sum_Y
    \sum_{\lambda=\pm1}\sqrt{(\ell+\lambda s_{Y})(\ell-\lambda s_{Y}+1)}\\\nonumber
    &&\qquad\,\times\,\int d\hn\,\bigg\{{}_{\lambda-s_{Y}}Y^*_{\ell m}(\hn)\left(C_{\ell}^{XY}-iC_\ell^{X\bar{Y}}\right){}_{s_Y}F^{(2),Y}_\lambda[x,y](\hn)\\\nonumber
    &&\qquad\qquad\,-\,{}_{s_Y-\lambda}Y^*_{\ell m}(\hn)\left(C_{\ell}^{XY}+iC_{\ell}^{X\bar{Y}}\right){}_{s_Y}F^{(2),Y*}_\lambda[x,y](\hn)
    \bigg\}\\\nonumber
    &&\,+\,\frac{1}{4}\sqrt{\ell(\ell+1)}C_\ell^{X\phi}\int d\hn\,\bigg\{{}_{\lambda} Y^*_{\ell m}(\hn)F^{(3)}_{\lambda}[x,y](\hn)-{}_{-\lambda} Y^*_{\ell m}(\hn)F^{(3)*}_{\lambda}[x,y](\hn)\bigg\}\,+\,(x\leftrightarrow y),
\eeq
defining the spin ($s_X,\lambda-s_Y,\lambda$) maps
\beq
    {}_{s_X}F^{(1),X}[x,y](\hn) &=& \sum_{\lambda=\pm1}{}_{s_X}V_\lambda^{{\rm lens},X}[x](\hn)V_{-\lambda}^{\rm ISW}[y](\hn)\\\nonumber
    {}_{s_Y}F^{(2),Y}_\lambda[x,y](\hn) &=& {}_{-s_{Y}}U^{Y}[x](\hn) V_{-\lambda}^{\rm ISW}[y](\hn)\\\nonumber
    F^{(3)}_\lambda[x,y](\hn) &=&\sum_Y{}_{s_Y}V_{-\lambda}^{{\rm lens},Y}[x](\hn){}_{-s_Y}U^{Y}[y](\hn).
\eeq
Notably, each term involves a complex-conjugate pair, such that $Q_{\ell m}^X$ is spin-zero. The full expression can be computed using spherical harmonic-transforms (up to spin-three), as before.

\subsection{Trispectrum Estimator}\label{app: tspec-pol}
\noindent As in \S\ref{subsec: tspec-estimator}, we can build quadratic estimators for the lensing potential, $\phi_{LM}$, and the unlensed fields, $a_{LM}^{Y}(\chi)$, by projecting \eqref{eq: lens-isw-quad-correlators-pol} onto two copies of the filtered data:
\beq\label{eq: lens-isw-quad-estimators-pol}
    \Phi^{\rm lens}_{LM}[x,y] &=& \frac{1}{\sqrt{L(L+1)}}\sum_{\ell_1\ell_3m_1m_3X_1X_3Y}\mathcal{I}^{\ell_1m_1X_1}_{\ell_3(-m_3)YL(-M)}C_{\ell_3}^{X_3Y}x^{X_1*}_{\ell_1m_1}y^{X_3*}_{\ell_3m_3}\\\nonumber
    &=& -\frac{1}{4}\sum_{Y}\sum_{\lambda=\pm1}\bigg\{\int d\hn\,\left[{}_{s_{Y}}U^{Y}[x](\hn){}_{s_{Y}}V_\lambda^{{\rm lens},Y*}[y](\hn)-{}_{s_{Y}}U^{Y*}[x](\hn){}_{s_{Y}}V^{{\rm lens},Y}_{-\lambda}[y](\hn)\right]{}_{+\lambda}Y^*_{LM}(\hn)\bigg\}\\\nonumber
    \Phi^{{\rm ISW},Y}_{LM}[x,y](\chi) &=& \frac{1}{\sqrt{L(L+1)}}\sum_{\ell_1\ell_3m_1m_3X_1X_3}\mathcal{I}^{\ell_1m_1X_1}_{L(-M)Y\ell_3(-m_3)}C_{\ell_3}^{X_3\phi}(\chi)x^{X_1*}_{\ell_1m_1}y^{X_3*}_{\ell_3m_3}\\\nonumber
    &=& \frac{1}{4}\sum_{\lambda=\pm1}\sqrt{\frac{(L+\lambda s_Y)(L-\lambda s_Y+1)}{L(L+1)}}\int d\hn\,
    \bigg[\left({}_{s_Y}U^Y[x](\hn)-i{}_{s_Y}U^{\bar{Y}}[x](\hn)\right)V_{\lambda}^{\rm ISW}[y](\chi){}_{s_Y-\lambda}Y^*_{LM}(\hn)\\\nonumber
    &&\quad\qquad\qquad\qquad\qquad\qquad\qquad\qquad\,+\,\left({}_{-s_Y}U^Y[x](\hn)+i{}_{-s_Y}U^{\bar{Y}}[x](\hn)\right)V_{-\lambda}^{\rm ISW}[y](\chi){}_{\lambda-s_Y}Y^*_{LM}(\hn)\bigg].
\eeq
These reduce to \eqref{eq: lens-isw-quad-estimators} for $X_i=Y=T$, and satisfy $(-1)\Phi^{*}_{L-M} = \Phi_{LM}$. Due to the addition of polarization indices, the ISW estimator takes a more complex form than the lensing equivalent, and additionally involves mixing of $E$- and $B$-modes.

Using the above definitions, we can form the full estimator numerators as in \eqref{eq: estimator-n=4}. For lensing, the result matches the temperature-only estimator of \eqref{eq: lens-tspec-estimator-T} since the polarization summation is contained within \eqref{eq: lens-isw-quad-estimators-pol}; for the ISW exchange shapes, we find the estimators:
\beq
    \widehat{\mathcal{N}}_{A_{\rm ISW-I}}[\alpha,\beta,\gamma,\delta] &=& \frac{1}{24}\sum_{LMYY'}(-1)^ML(L+1)\Phi^{{\rm ISW},Y}_{LM}[\Si\alpha,\Si\beta]\Phi^{{\rm ISW},Y'}_{L(-M)}[\Si\gamma,\Si\delta]C_L^{YY'}+\text{11 perms.}\nonumber\\
    \widehat{\mathcal{N}}_{A_{\rm ISW-II+III}}[\alpha,\beta,\gamma,\delta] &=& \frac{1}{24}\sum_{LMY}(-1)^ML(L+1)\int_0^{\chi_\star}d\chi\,C_L^{\phi Y}(\chi)\\\nonumber
    &&\,\times\,\bigg(\Phi^{{\rm ISW},Y}_{LM}[\Si\alpha,\Si\beta](\chi)\Phi^{{\rm lens}*}_{L(-M)}[\Si\gamma,\Si\delta]\\\nonumber
    &&\qquad\,+\,\Phi^{\rm lens}_{LM}[\Si\alpha,\Si\beta]\Phi^{{\rm ISW},Y*}_{LM}[\Si\gamma,\Si\delta](\chi)\bigg)+\text{11 perms.},
\eeq
which are simply the weighted auto- and cross-spectra of \eqref{eq: lens-isw-quad-estimators-pol}, summed over polarizations and simplifying the $\chi$ integrals where possible. Similarly, the contact ISW estimator has the numerator
\beq
    \widehat{\mathcal{N}}_{A_{\rm ISW-IV}}[\alpha,\beta,\gamma,\delta]&=& \frac{1}{192}\sum_Y\sum_{\lambda=\pm1}\int d\hn\,\bigg[{}_{-s_Y}U^{Y}[\Si\alpha](\hn)V_{-\lambda}^{\rm ISW}[\Si\gamma](\hn)\\\nonumber
    &&\,\times\,\bigg\{{}_{s_Y}G^{1,Y}_{\lambda}[\Si\beta](\hn)V_{-\lambda}^{\rm ISW}[\Si\delta](\hn)+{}_{s_Y}G^{0,Y}_{\lambda}[\Si\beta](\hn)V_{\lambda}^{\rm ISW}[\Si\delta](\hn)\bigg\}+\text{c.c.}\bigg]\,+\,\text{11 perms.}
\eeq
where we have defined the spin-$(s-2n\lambda)$ field
\beq
    {}_{s}G^{n,Y}_\lambda[x](\hn)=\sum_{\ell mZ}\sqrt{(\ell+\lambda s)(\ell-\lambda s+1)(\ell+\lambda s-n)(\ell-\lambda s+1+n)}\left[C_{\ell}^{ZY}-iC_{\ell}^{Z\bar{Y}}\right]{}_{s-2n\lambda}Y_{\ell m}(\hn)x_{\ell m}^Z.
\eeq
Noting that ${}_0G^{|\lambda|,T}_\lambda$ is equal to the $S^{\rm lens}_\lambda$ function defined in \eqref{eq: S-lens-def}, this recovers \eqref{eq: isw-con-tspec-estimator-T} in the temperature-only limit.

In addition to the numerators, we require the Fisher matrix normalization, $\F_{\alpha\beta}$. As in \S\ref{subsec: tspec-estimator}, this can be computed via Monte Carlo summation, generalizing \eqref{eq: fish4} by summing over polarization states. This requires $Q_{\ell m,\alpha}^{X}$ functions for each template of interest -- these are presented in Appendix \ref{app: Q-derivs} and can be computed using spin-weighted spherical harmonic transforms.

\section{Fisher Derivatives}\label{app: Q-derivs}
\noindent Below, we list the $Q_{\ell m}$ derivatives used in the lensing and ISW-lensing trispectrum estimators. In all cases, we omit the lengthy, though elementary, derivations. First, we give the temperature-only results used in \S\ref{subsec: tspec-estimator}, before presenting the extension to polarization, as in Appendix \ref{app: tspec-pol}.

\subsection{Temperature}
\noindent The lensing $Q_{\ell m}$ function can be obtained from combining \eqref{eq: Q-deriv}\,\&\,\eqref{eq: tris-ex-full-sky-T}, which yields
\beq
    Q_{\ell m,A_{\rm lens}}[x,y,z] &=& \sum_{\ell_3m_3LM}(-1)^{M}\left(\mathcal{I}^{\ell m}_{LM\ell_3(-m_3)}C_{\ell_3}^{TT}+\mathcal{I}^{\ell_3m_3}_{LM\ell(-m)}C_\ell^{TT}\right)\sqrt{L(L+1)}\Phi_{LM}^{\rm lens}[x,z]y^*_{\ell_3m_3}C_L^{\phi\phi}+\text{5 perms.}\nonumber\\
    &=&-\frac{1}{2}\sum_{\lambda=\pm1}\int d\hn\,W^{\rm lens-lens}_{\lambda}[x,z](\hn)\\\nonumber
    &&\qquad\,\times\,\left(Y^*_{\ell m}(\hn)V_\lambda^{\rm lens}[y](\hn)-\sqrt{\ell(\ell+1)}C_\ell^{TT}{}_{\lambda}Y^*_{\ell m}(\hn)U[y](\hn)\right)+\text{5 perms.}
\eeq
matching \citep{Philcox4pt1}, keeping track of asymmetric permutations and defining
\beq
    W^{\rm lens-lens}_{\lambda}[x,y](\hn) = \sum_{LM}{}_{\lambda}Y_{LM}(\hn)L(L+1)\Phi_{LM}^{\rm lens}[x,y]C_L^{\phi\phi}.
\eeq 
The exchange ISW-lensing contributions (ISW-I and ISW-II/ISW-III) are analogous:
\beq\label{eq: exchange-Qlm-T}
    Q_{\ell m,A_{\rm ISW-I}}[x,y,z] 
    %
    &=&-\frac{1}{2}\sum_{\lambda=\pm1}\int d\hn\,W^{\rm ISW-ISW}_{\lambda}[x,z](\hn)\nonumber\\
    &&\qquad\,\times\,\left(Y^*_{\ell m}(\hn)V_\lambda^{\rm ISW}[y](\hn)-\sqrt{\ell(\ell+1)}C_\ell^{T\phi}{}_{\lambda}Y^*_{\ell m}(\hn)U[y](\hn)\right)+\text{5 perms.}\\\nonumber
    Q_{\ell m,A_{\rm ISW-II+III}}[x,y,z] &=&-\frac{1}{2}\sum_{\lambda=\pm1}\int d\hn\,W^{\rm lens-ISW}_{\lambda}[x,z](\hn)\\\nonumber
    &&\qquad\,\times\,\left(Y^*_{\ell m}(\hn)V_\lambda^{\rm lens}[y](\hn)-\sqrt{\ell(\ell+1)}C_\ell^{TT}{}_{\lambda}Y^*_{\ell m}(\hn)U[y](\hn)\right)+\text{5 perms.}\\\nonumber
    &&-\frac{1}{2}\sum_{\lambda=\pm1}\int_0^{\chi_\star}d\chi\int d\hn\,W^{\rm ISW-lens}_{\lambda}[x,z](\hn,\chi)\\\nonumber
    &&\qquad\,\times\,\left(Y^*_{\ell m}(\hn)V_\lambda^{\rm ISW}[y](\hn,\chi)-\sqrt{\ell(\ell+1)}C_\ell^{T\phi}(\chi){}_{\lambda}Y^*_{\ell m}(\hn)U[y](\hn)\right)+\text{5 perms.}
\eeq
with the definitions
\beq
    W^{\rm ISW-ISW}_{\lambda}[x,y](\hn) &=& \sum_{LM}{}_{\lambda}Y_{LM}(\hn)L(L+1)\Phi_{LM}^{\rm ISW}[x,y]C_L^{TT}\\\nonumber
    W^{\rm lens-ISW}_{\lambda}[x,y](\hn) &=& \sum_{LM}{}_{\lambda}Y_{LM}(\hn)L(L+1)\int_0^{\chi_\star} d\chi\,\Phi_{LM}^{\rm ISW}[x,y](\chi)C_L^{\phi T}(\chi)\\\nonumber
    W^{\rm ISW-lens}_{\lambda}[x,y](\hn,\chi) &=& \sum_{LM}{}_{\lambda}Y_{LM}(\hn)L(L+1)\Phi_{LM}^{\rm lens}[x,y]C_L^{\phi T}(\chi).
\eeq
As for the numerators, the ISW and lensing functions are equivalent upon exchange of $C_\ell^{TT}\to C_\ell^{T\phi}$ and $C_L^{\phi\phi}\to C_L^{TT}$. In practice, we simplify the ISW-II/ISW-III definitions by evaluating the integral at $\chi=\chi_{\rm eff}$, with $C_L^{\phi T}(\chi)\to C_L^{T\phi}$, as discussed in \S\ref{subsec: tspec-estimator}.

For the contact term (ISW-IV), we start from \eqref{eq: tris-con-full-sky-T}, finding
\beq
    Q_{\ell m,A_{\rm ISW-IV}}[x,y,z]&=&\frac{1}{8}\sum_{\lambda=\pm1}\int d\hn\,Y^*_{\ell m}(\hn)V_{\lambda}^{\rm ISW}[z](\hn)\\\nonumber
    &&\qquad\,\times\,\bigg[S_0^{\rm lens}[x](\hn)V_{-\lambda}^{\rm ISW}[y](\hn)+S_{\lambda}^{\rm lens}[x](\hn)V^{\rm ISW}_{\lambda}[y](\hn)\bigg]\\\nonumber
    &+&\frac{1}{8}\sqrt{\ell(\ell+1)}C^{TT}_{\ell}\sum_{\lambda=\pm1}\int d\hn\,U[x](\hn)V_{\lambda}^{\rm ISW}[z](\hn)\\\nonumber
    &&\qquad\,\times\,\bigg[\sqrt{\ell(\ell+1)}V^{\rm ISW}_{-\lambda}[y](\hn)Y^*_{\ell m}(\hn)+\sqrt{(\ell-1)(\ell+2)}V^{\rm ISW}_{\lambda}[y](\hn){}_{-2\lambda}Y^*_{\ell m}(\hn)\bigg]\\\nonumber
    &-&\frac{1}{4}\sqrt{\ell(\ell+1)}C_\ell^{T\phi}\sum_{\lambda=\pm1}\int d\hn\,U[y](\hn)V_{\lambda}^{\rm ISW}[z](\hn)\\\nonumber
    &&\qquad\,\times\,\bigg[S_0^{\rm lens}[x](\hn){}_{-\lambda}Y^*_{\ell m}(\hn)+S_{\lambda}^{\rm lens}[x](\hn){}_{\lambda}Y^*_{\ell m}(\hn)\bigg]\\\nonumber
    &+&\text{5 perms.}
    %
\eeq
This, and all other derivatives, can be computed using spherical harmonic transforms (up to spin-two).

\subsection{Polarization}
\noindent Computation of the polarized trispectrum estimator normalizations proceeds analogously to the temperature-only limit. Due to the additional polarization indices, the resulting expressions are fairly gargantuan but can be efficiently implemented using spherical-harmonic transforms. For lensing, substituting \eqref{eq: tris-ex-full-sky-pol} into \eqref{eq: Q-deriv} and simplifying leads to
\beq\label{eq: Q-lens-pol}
    Q_{\ell m,A_{\rm lens}}^{X}[x,y,z] &=&\frac{1}{4}\sum_{\lambda=\pm1}\int d\hn\,\bigg[{}_{s_X}V_{-\lambda}^{{\rm lens},X*}[y](\hn)W^{\rm lens-lens}_{\lambda}[x,z](\hn){}_{-s_X}Y_{\ell m}^*(\hn)\\\nonumber
    &&\qquad\qquad\qquad\quad\,+\,{}_{s_X}V_{-\lambda}^{{\rm lens},X}[y](\hn)W_{\lambda}^{\rm lens-lens*}[x,z](\hn){}_{s_X}Y_{\ell m}^*(\hn)\bigg]\\\nonumber
    &&\,+\,\frac{1}{4}\sum_{\lambda=\pm1}\sum_Y\sqrt{(\ell+\lambda s_Y)(\ell-\lambda s_Y+1)}\\\nonumber
    &&\,\times\,\bigg\{\left(C_{\ell}^{YX}-iC_\ell^{\bar{Y}X}\right)\int d\hn\,{}_{-s_{Y}}U^{Y}[y](\hn)W^{\rm lens-lens}_\lambda[x,z](\hn){}_{-s_Y+\lambda}Y_{\ell m}^*(\hn)\\\nonumber
    &&\qquad\,-\,\left(C_{\ell}^{YX}+iC_\ell^{\bar{Y}X}\right)\int d\hn\,{}_{-s_{Y}}U^{Y*}[y](\hn)W^{\rm lens-lens*}_{\lambda}[x,z](\hn){}_{s_Y-\lambda}Y_{\ell m}^*(\hn)\bigg\}\,+\,\text{5 perms.},
\eeq
matching \citep{Philcox4pt1}, defining the spin-$\lambda$ map
\beq
   W^{{\rm lens-lens}}_\lambda[x,y](\hn) &\equiv& \sum_{LMY}{}_{\lambda}Y_{LM}(\hn)L(L+1)\Phi^{{\rm lens}}_{LM}[x,y]C_L^{\phi\phi}
\eeq
with $W_{-\lambda} = (-1)^\lambda W_\lambda^*$. Similarly, the exchange ISW-lensing functions are given by
\beq
    Q^X_{\ell m,A_{\rm ISW-I}}[x,y,z] &=&\frac{1}{4}\sum_{\lambda=\pm1}\int d\hn\,\bigg[V_{-\lambda}^{{\rm ISW}*}[y](\hn){}_{s_X}W_\lambda^{{\rm ISW-ISW},X}[x,z](\hn){}_{-s_X}Y^*_{\ell m}(\hn) \\\nonumber
    &&\qquad\qquad\qquad\,+\, V_{-\lambda}^{{\rm ISW}}[y](\hn){}_{s_X}W^{{\rm ISW-ISW},X*}_\lambda[x,z](\hn){}_{s_X}Y^*_{\ell m}(\hn)\bigg]\\\nonumber
    &&\,+\,\frac{1}{4}\sqrt{\ell(\ell+1)}C_{\ell}^{X\phi}\sum_{Y}\sum_{\lambda=\pm1}\int d\hn\,\bigg[{}_{s_Y}U^{Y}[y](\hn){}_{s_Y}W^{{\rm ISW-ISW},Y}_{\lambda}[x,z](\hn){}_\lambda Y^*_{\ell m}(\hn)\\\nonumber
    &&\qquad\qquad\qquad\qquad\qquad\qquad\qquad\qquad\,-\,{}_{s_Y}U^{Y*}[y](\hn){}_{s_Y}W^{{\rm ISW-ISW},Y*}_{\lambda}[x,z](\hn){}_{-\lambda} Y^*_{\ell m}(\hn)\bigg] \\\nonumber
    &&\,+\,\text{5 perms.}
\eeq
and
\beq
    Q^X_{\ell m,A_{\rm ISW-II+III}}[x,y,z] &=&\frac{1}{4}\sum_{\lambda=\pm1}\int_0^{\chi_\star}d\chi\,\int d\hn\,\bigg[V_{-\lambda}^{\rm ISW*}[y](\hn,\chi){}_{s_X}W^{{\rm ISW-lens},X}_\lambda[x,z](\hn,\chi){}_{-s_X}Y^*_{\ell m}(\hn)\\\nonumber
    &&\qquad\qquad\qquad\,+\,V_{-\lambda}^{\rm ISW}[y](\hn,\chi){}_{s_X}W^{{\rm ISW-lens},X*}_\lambda[x,z](\hn,\chi){}_{s_X}Y^*_{\ell m}(\hn)\bigg]\\\nonumber
    &&\,+\,
    \frac{1}{4}\sqrt{\ell(\ell+1)}\int_0^{\chi_\star}d\chi\,C_{\ell}^{X\phi}(\chi)\sum_{Y}\sum_{\lambda=\pm1}\int d\hn\,\bigg[{}_{s_Y}U^{Y}[y](\hn){}_{s_Y}W_\lambda^{{\rm ISW-lens},Y}[x,z](\hn,\chi) {}_{\lambda}Y^*_{\ell m}(\hn)\\\nonumber
    &&\qquad\qquad\qquad\qquad\qquad\qquad\qquad\,-\,{}_{s_Y}U^{Y*}[y](\hn){}_{s_Y}W_\lambda^{{\rm ISW-lens},Y*}[x,z](\hn,\chi) {}_{-\lambda}Y^*_{\ell m}(\hn)\bigg]\\\nonumber
    &&\,+\,\frac{1}{4}\sum_{\lambda=\pm1}\int d\hn\,\bigg[ {}_{s_X}V^{X,\rm lens*}_{-\lambda}[y](
    \hn)W^{{\rm lens-ISW},X}_{\lambda}[x,z](\hn){}_{-s_X}Y^*_{\ell m}(\hn)\\\nonumber
    &&\qquad\qquad\qquad\qquad+{}_{s_X}V^{{\rm lens},X}_{-\lambda}[y](
    \hn)W^{{\rm lens-ISW},X*}_{\lambda}[x,z](\hn){}_{s_X}Y^*_{\ell m}(\hn) \bigg]\\\nonumber
    &&\,+\,\frac{1}{4}\sum_{\lambda=\pm1}\sum_Y\sqrt{(\ell+\lambda s_Y)(\ell-\lambda s_Y+1)}\\\nonumber
    &&\qquad\qquad\qquad\bigg\{\left(C_{\ell}^{XY}-iC_{\ell}^{X\bar{Y}}\right)\int d\hn\,{}_{-s_{Y}}U^{Y}[y](\hn)W_{\lambda}^{{\rm lens-ISW}}[x,z](\hn){}_{-s_{Y}+\lambda}Y^*_{\ell m}(\hn)\\\nonumber
    &&\qquad\qquad\qquad\qquad\,-\,\left(C_{\ell}^{XY}+iC_{\ell}^{X\bar{Y}}\right)\int d\hn\,{}_{-s_{Y}}U^{Y*}[y](\hn)W_{\lambda}^{{\rm lens-ISW}*}[x,z](\hn){}_{s_{Y}-\lambda}Y^*_{\ell m}(\hn) \bigg\}\\\nonumber
    &&\,+\,\text{5 perms.}
\eeq
defining
\beq
    {}_{s_X}W^{{\rm ISW-ISW},X}_\lambda[x,y](\hn) &=& \sum_{LMY}{}_{-s_X+\lambda}Y_{LM}(\hn)\sqrt{L(L+1)(L+\lambda s_X)(L-\lambda s_X+1)}\Phi^{{\rm ISW},Y}_{LM}[x,y]\left(C_L^{XY}+iC_L^{\bar{X}Y}\right)\nonumber\\
    W^{{\rm lens-ISW}}_{\lambda}[x,y](\hn) &=& \sum_{LMY'}{}_{\lambda}Y_{LM}(\hn)L(L+1)\int_0^{\chi_\star}d\chi\,\Phi^{{\rm ISW},Y'}_{LM}[x,y](\chi)C_L^{\phi Y'}(\chi)\\\nonumber
    {}_{s_X}W^{{\rm ISW-lens},X}_\lambda[x,y](\hn,\chi) &=& \sum_{LMY}{}_{-s_X+\lambda}Y_{LM}(\hn)\sqrt{L(L+1)(L+\lambda s_X)(L-\lambda s_X+1)}\Phi^{{\rm lens}}_{LM}[x,y]\left(C_L^{\phi X}(\chi)+iC_L^{\phi\bar{X}}(\chi)\right).
\eeq
These simplify to \eqref{eq: exchange-Qlm-T} in the temperature-only limit.

The contact $Q_{\ell m}^X$ function can be obtained from \eqref{eq: tris-con-full-sky-pol}, and takes the form 
\beq
    Q^X_{\ell m,A_{\rm ISW-IV}}[x,y,z] 
    &=& \frac{1}{16}\int d\hn\,\left[{}_{s_X}Y^*_{\ell m}(\hn){}_{s_X}F^{(1),X}[x,y,z](\hn)+{}_{-s_X}Y^*_{\ell m}(\hn){}_{s_X}F^{(1),X*}[x,y,z](\hn)\right]\\\nonumber
    &&\,+\,\frac{1}{16}\sum_Y\sum_{\lambda=\pm1}\sqrt{(\ell+\lambda s_Y)(\ell-\lambda s_Y+1)(\ell+\lambda s_Y-1)(\ell-\lambda s_Y+2)}\\\nonumber
    &&\,\times\,\int d\hn\,\bigg[\left(C_{\ell}^{XY}-iC_{\ell}^{X\bar{Y}}\right){}_{s_Y}F^{(2),Y}_\lambda[x,y,z](\hn){}_{-s_Y+2\lambda}Y^*_{\ell m}(\hn)\\\nonumber
    &&\qquad\qquad\,+\,\left(C_{\ell}^{XY}+iC_{\ell}^{X\bar{Y}}\right){}_{s_Y}F^{(2),Y*}_\lambda[x,y,z](\hn){}_{s_Y-2\lambda}Y^*_{\ell m}(\hn)\bigg]\\\nonumber
    &&\,+\,\frac{1}{16}\sum_Y\sum_{\lambda=\pm1}(\ell+\lambda s_Y)(\ell-\lambda s_Y+1)\\\nonumber
    &&\,\times\,\int d\hn\,\bigg[\left(C_{\ell}^{XY}-iC_{\ell}^{X\bar{Y}}\right){}_{s_Y}F^{(3),Y}_\lambda[x,y,z](\hn){}_{-s_Y}Y^*_{\ell m}(\hn)\\\nonumber
    &&\qquad\qquad\,+\,\left(C_{\ell}^{XY}+iC_{\ell}^{X\bar{Y}}\right){}_{s_Y}F^{(3),Y*}_\lambda[x,y,z](\hn){}_{s_Y}Y^*_{\ell m}(\hn)\bigg]\\\nonumber
    &&\,-\,\frac{1}{8}\sqrt{\ell(\ell+1)}C_{\ell}^{X\phi}\sum_{\lambda=\pm1}\int d\hn\,\bigg[F^{(4)}_\lambda[x,y,z](\hn){}_{\lambda}Y^*_{\ell m}(\hn) - F^{(4)*}_\lambda[x,y,z](\hn){}_{-\lambda}Y^*_{\ell m}(\hn)\bigg]\\\nonumber
    &&\,+\,\text{5 perms.}
\eeq
where each term is the sum of a positive and a negative spin-map:
\beq
    {}_{s_X}F^{(1),X}[x,y,z](\hn) &=& \sum_{\lambda=\pm1}V_{-\lambda}^{{\rm ISW}}[y](\hn)\bigg\{{}_{s_X}G^{1,X}_\lambda[x](\hn)V_{-\lambda}^{{\rm ISW}}[z](\hn)\,+\,{}_{s_X}G^{0,X}_\lambda[x](\hn)V_{\lambda}^{{\rm ISW}}[z](\hn)\bigg\}\\\nonumber
    {}_{s_Y}F^{(2),Y}_\lambda[x,y,z](\hn) &=& {}_{-s_Y}U^{Y}[x](\hn)V_{-\lambda}^{\rm ISW}[y](\hn)V_{-\lambda}^{\rm ISW}[z](\hn)\\\nonumber
    {}_{s_Y}F^{(3),Y}_\lambda[x,y,z](\hn) &=& {}_{-s_Y}U^{Y}[x](\hn)V_{-\lambda}^{\rm ISW}[y](\hn)V_{\lambda}^{\rm ISW}[z](\hn)\\\nonumber
    F_\lambda^{(4)}[x,y,z](\hn) &=& \sum_Y{}_{-s_Y}U^{Y}[y](\hn)\left(V_{-\lambda}^{\rm ISW}[z](\hn){}_{s_Y}G_\lambda^{1,Y}[x](\hn)\,+\,V_{\lambda}^{\rm ISW}[z](\hn){}_{s_Y}G_\lambda^{0,Y}[x](\hn)\right).
\eeq
This can be estimated via spherical-harmonic-transforms, as before.

\bibliographystyle{apsrev4-1}
\bibliography{refs}

\end{document}